\documentclass[sigconf]{acmart}

\usepackage{amsmath,amsfonts}
\usepackage{mathtools}

\usepackage{pifont} %use the command \ding

\usepackage{bbding} %对勾叉号
\usepackage{pifont}

\usepackage{makecell}
\usepackage{colortbl} % table color

\usepackage{tikz}
\usetikzlibrary{arrows.meta}
\usetikzlibrary{automata,positioning}

\usepackage{threeparttable}

\newcommand{\ethsym}{%
  {\textcolor{blue!60}{\rotatebox{-8}{\scalebox{1.3}{$\blacklozenge$}}}}%
}

\newcommand{\ercsym}{%
  {\textcolor{magenta!70!red}{\rotatebox{-8}{\scalebox{1.3}{$\blacklozenge$}}}}%
}

\renewcommand*{\arraystretch}{1.5}%
\definecolor{tabred}{RGB}{230,36,0}%
\definecolor{tabgreen}{RGB}{0,116,21}%
\definecolor{taborange}{RGB}{250,124,30}%
\definecolor{tabbrown}{RGB}{171,70,0}%
\definecolor{tabyellow}{RGB}{251,253,169}%
\newcommand*{\vcorr}{%
  \vadjust{\vspace{-\dp\csname @arstrutbox\endcsname}}%
  \global\let\vcorr\relax
}% 

\usepackage{mathtools}

\usepackage{lscape}
\usepackage[figuresright]{rotating}
\usepackage[graphicx]{realboxes}
\usepackage{adjustbox}
\usepackage{caption}

\usepackage{subfigure}

\usepackage{colortbl} 
\usepackage{xfrac}

\usepackage{appendix}
\usepackage{float}

\usepackage{fbox} %use box
\usepackage{fancybox}%use other types of boxes
\usepackage{framed}

\usepackage{multirow}

\def\BibTeX{{\rm B\kern-.05em{\sc i\kern-.025em b}\kern-.08em
    T\kern-.1667em\lower.7ex\hbox{E}\kern-.125emX}}
    
\usepackage{CJKutf8}  
\usepackage{pgfplots}
\usepackage{filecontents}

\usepackage{hyperref}

\usepackage{tabularx,makecell, caption}
\usepackage{enumitem} % to control Itemization spacing

\newcolumntype{L}{>{\arraybackslash}X}

\usepackage{multicol}
\usepackage{listings}
\usepackage{blindtext}
\usepackage{etoolbox,xstring,mfirstuc,textcase}
\usepackage{listings}

\usepackage{subfiles}
\usepackage[most]{tcolorbox}

\usepackage{xcolor}
\usepackage{etoolbox}
\AfterEndEnvironment{lstlisting}{\vspace{0.75em}}
\makeatletter
\pretocmd{\lst@MakeCaption}{\vspace{0.8em}}{}{}
\makeatother
\usepackage{listings}
\lstdefinelanguage{text}{
    alsoletter={},
    morestring=[b]",
}

\theoremstyle{plain}

\theoremstyle{definition}

\usepackage{siunitx}
\usepackage{comment}
\usepackage{indentfirst} 
\usepackage{framed} 

\usepackage[font=small,labelfont=bf,tableposition=top]{caption}
\usepackage{booktabs}
\usepackage{dashbox}

\usepackage{listings}
\usepackage{url}

\newenvironment{packeditemize}{
	\begin{list}{$\bullet$}{
			\setlength{\labelwidth}{4pt}
			\setlength{\itemsep}{0pt}
			\setlength{\leftmargin}{\labelwidth}
			\addtolength{\leftmargin}{\labelsep}
			\setlength{\parindent}{0pt}
			\setlength{\listparindent}{\parindent}
			\setlength{\parsep}{0pt}
			\setlength{\topsep}{1pt}}}{\end{list}}

\lstdefinelanguage{Javascript}{
  morekeywords={
    break, case, catch, continue, debugger, default, delete, do, else,
    finally, for, function, if, in, instanceof, new, return, switch,
    this, throw, try, typeof, var, void, while, with, const, let
  },
  sensitive=true,
  morecomment=[l]{//},
  morecomment=[s]{/*}{*/},
  morestring=[b]',
  morestring=[b]"
}

\usepackage{color}
\usepackage{algorithm,algpseudocode} %% select one of them

\renewcommand\footnotetextcopyrightpermission[1]{} % removes footnote with conference information in first column

\begin{document}
\title{EIP-7702 Phishing Attack}

%=================================================
%author
%=================================================

%\begin{comment} 

\author{Minfeng Qi$^{1}$, Qin Wang$^{2}$, Ruiqiang Li$^{3}$, Tianqing Zhu$^{1}$, Shiping Chen$^{2}$}
%\thanks{$^{\textcolor{green}{*}}$ Equal contribution}%Equal contribution} effective when adding additional authors
\medskip
\affiliation{
\textit{$^1$City University of Macau} $|$ \textit{$^2$CSIRO Data61}   $|$ \textit{$^3$University of Wollongong}
}

%\end{comment}

%=================================================
%abstract
%=================================================

\begin{abstract}
EIP-7702 introduces a delegation-based authorization mechanism that allows an externally owned account (EOA) to authenticate a single authorization tuple, after which all subsequent calls are routed to arbitrary delegate code. We show that this design enables a qualitatively new class of phishing attacks: \textit{instead of deceiving users into signing individual transactions, an attacker can induce a victim to sign a single authorization tuple that grants unconditional and persistent execution control over the account.}

Through controlled experiments, we identify three reliable trigger pathways: user-driven, attacker-driven, and protocol-triggered. Each can lead to full account takeover and complete asset drainage. We further propose two extended attack surfaces. First, ERC-4337’s \textsf{EntryPoint} pipeline enables remote and repeated activation of the delegated code without further victim involvement. Second, the chain-agnostic authorization mode permits replay-like compromises across independent networks. 

We also present the first empirical measurement of EIP-7702 usage across major EVM chains. Analyzing over 150k authorization and execution events involving 26k addresses and hundreds of delegator contracts, we assess the protocol’s real-world footprint. Our findings show that EIP-7702 authorizations are highly centralized, dominated by a small number of contract families linked to criminal activity and repeatedly reused across incidents. Corresponding loss data reveals substantial theft of ETH, ERC-20 tokens, and NFTs. These results provide practical evidence that the attack surface we identify is not merely theoretical, but is already being exploited at scale. We conclude by proposing protocol-level defenses to mitigate the delegation-based phishing vector introduced by EIP-7702.
\end{abstract}


%=================================================
\maketitle
%=================================================   

%=================================================   
\section{Introduction}
%=================================================   

Phishing remains one of the most persistent and damaging attack vectors in the cryptocurrency ecosystem. As users increasingly rely on smart-contract wallets, NFT marketplaces, DeFi routers, and cross-chain bridges, adversaries have shifted from traditional credential theft toward transaction-centric and semantics-aware attacks~\cite{wang2025tsgn,ghosh2025temper}. Recent studies (summarized in Table~\ref{tab:phishing_attack_comparison}) show that modern blockchain phishing leverages UI-level deception through fake DApps and wallet-connect prompts~\cite{he2023txphishscope,idehen2024web3}, malicious calldata and misleading payloads embedded in legitimate protocols~\cite{microsoft2022icephish,chainalysis2023approval,slowmist2023permit,veritas_permit}, and fallback or proxy redirection via adversarial contracts~\cite{yang2024stole,darkreading2024opensea}. Even DeFi routers and MEV bots have become targets of transaction-level manipulation~\cite{microsoft2022icephish,cryptonews2025defi,morales2025mev}. Yet Ethereum’s dominant account model remains the externally owned account (EOA)\footnote{EOA is a type of Ethereum account controlled by a private key, which enables its holder to send transactions, hold tokens, and interact with smart contracts~\cite{stackup_eoa}.}, whose validation logic is hard-wired and cannot enforce fine-grained execution policies. This mismatch between evolving phishing techniques and the static semantics of EOAs creates a structural security gap.

\begin{figure}[!]
\vspace{0.8em}
\centering
    \includegraphics[width=0.93\linewidth]{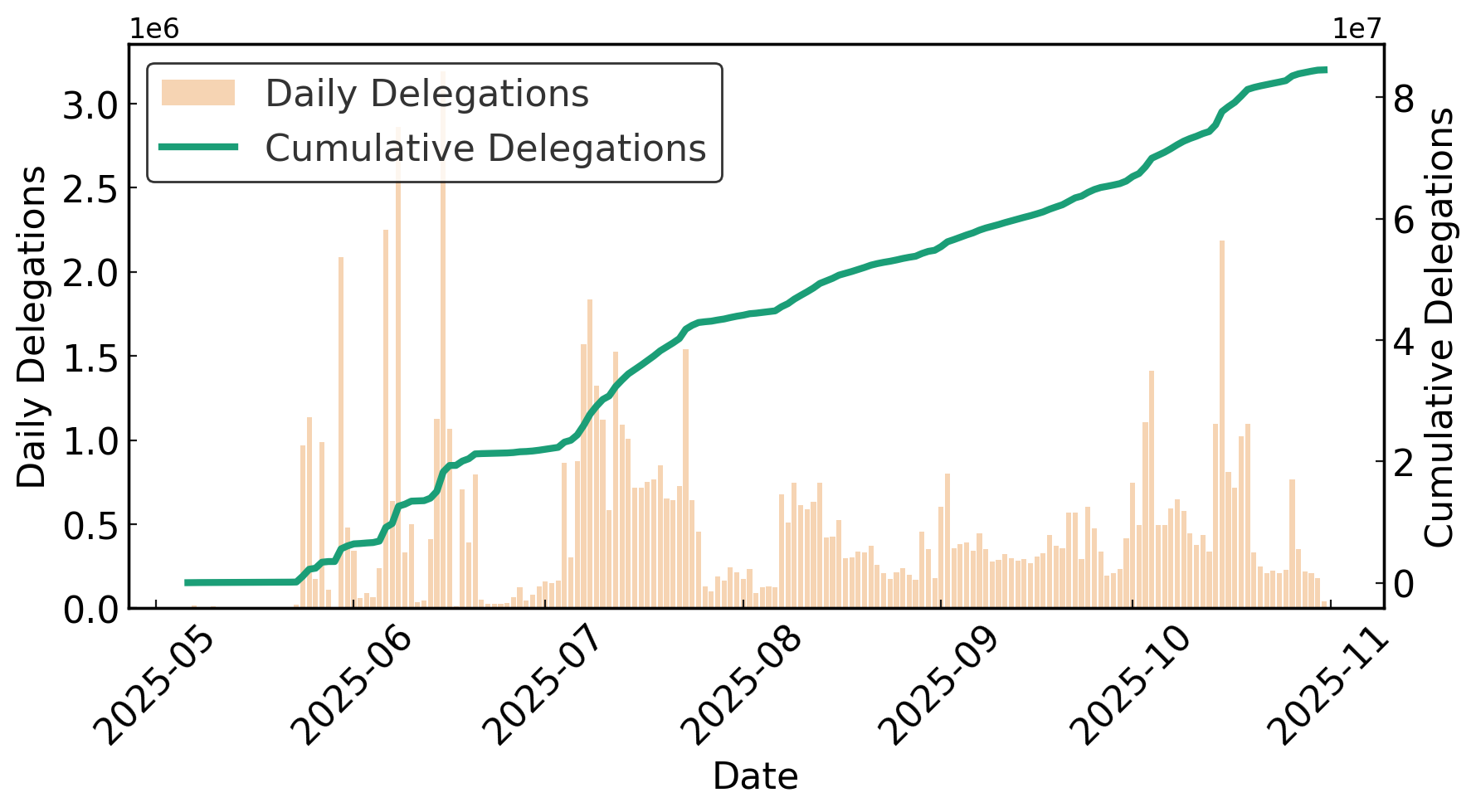}
     \vspace{-1em}
\caption{EIP-7702 Transaction Count and Growth Trend.}
\label{fig:growth_trend}
 \vspace{-1.5em}
\end{figure}

Account Abstraction (AA) was introduced to mitigate this gap by enabling programmable validation and flexible security rules~\cite{ethereum_account_abstraction}. Although ERC-4337~\cite{erc4337} offers AA without protocol changes, EOAs continue to dominate the ecosystem, leading to interim mechanisms that attempt to extend EOA capabilities. EIP-7702 is one such mechanism and has seen rapid adoption,with 2--3 million delegations per day observed on major EVM chains (Fig.~\ref{fig:growth_trend}). Instead of converting an EOA into a contract, EIP-7702 introduces a lightweight delegation pathway: a user signs an authorization tuple that temporarily overwrites the account’s code field with a delegating stub, after which all calls to the EOA are routed into a delegate contract.

\begin{table*}[!ht]
\centering
\caption{\textcolor{teal}{Phishing-related Attacks.} 
Phishing attacks in blockchain ecosystems span user deception, transaction-level payload manipulation, 
contract-level semantic abuse, and vertical-specific traps (NFT, DeFi, MEV). 
Phishing attacks manipulate user intent and signature flows, 
often combining UI deception, malicious calldata, proxy misuse, or approval hijacking to seize digital assets  (details are discussioned in \S\ref{sec-rw}).}
\label{tab:phishing_attack_comparison}
\vspace{-0.8em}

\renewcommand{\arraystretch}{1.1}
\resizebox{\textwidth}{!}{%
\begin{tabular}{c|cc|cc|c}

\midrule

\multicolumn{1}{c}{\textbf{Aspect}} &
\multicolumn{1}{c}{\textbf{Target}} &
\multicolumn{1}{c}{\textbf{Objective}} &
\multicolumn{1}{c}{\textbf{Key Exploit}} &
\multicolumn{1}{c}{\textbf{Mechanism}} &
\multicolumn{1}{c}{\textbf{Platform}} \\

\midrule
% ----------------------- Block 1: User-level deception -----------------------

Fake DApp / Website~\cite{he2023txphishscope,idehen2024web3} 
& User 
& Malicious signature 
& UI spoofing 
& Fake interfaces, wallet-connect traps 
& All \\

Airdrop / Giveaway Scam~\cite{li2023double,filiba2024scams,slowmist2025analysis} 
& User 
& Credential theft 
& Social engineering 
& Reward bait, phishing links 
& EVM-compatible \\

Impersonation Scam~\cite{chainalysis2023midyear} 
& User 
& Asset theft 
& Identity abuse 
& Fake admin/mod communication 
& All \\

\cmidrule{1-2}

% ----------------------- Block 2: Transaction-level phishing -----------------------

Payload-based Phishing~\cite{chen2024dissecting,pshanichnaya2024solana}
& Tx calldata 
& Token drain 
& Malicious payload 
& Crafted calldata misleading UI/signing 
& EVM-compatible \\

Approval Phishing~\cite{microsoft2022icephish,chainalysis2023approval} 
& Token approvals 
& Unlimited asset control 
& Infinite approval 
& Deceptive approval request 
& EVM-compatible \\

Simulation Deception~\cite{cholakov2025drainers} 
& Tx preview 
& Mask harmful effects 
& Simulation mismatch 
& Misleading view-call / gas griefing 
& EVM-compatible \\

\cmidrule{1-2}

% ----------------------- Block 3: Contract-level phishing -----------------------

Fake Mint / Claim Contracts~\cite{yang2024stole}
& NFT/tokens 
& NFT theft 
& Callback/fallback trap 
& Malicious mint/claim UX 
& NFT / EVM \\

Proxy-based Phishing~\cite{hacken2023honeypot}  
& Upgradeable contracts 
& Contract takeover 
& Proxy fallback abuse 
& Unauthorized logic override 
& EVM-compatible \\

Permit-based Phishing~\cite{slowmist2023permit,veritas_permit} 
& Typed signatures 
& Replay / infinite allowance 
& Signature abstraction 
& Trick user to sign typed-data 
& EVM-compatible \\

\cmidrule{1-2}

% ----------------------- Block 4: Ecosystem-specific phishing -----------------------

NFT Approval Scam~\cite{yang2024stole,darkreading2024opensea}
& NFT approvals 
& NFT seizure 
& setApprovalForAll misuse 
& Fake marketplace / cancel listing trap 
& NFT \\

DeFi Router Phishing~\cite{microsoft2022icephish,cryptonews2025defi}  
& Swap/LP ops 
& LP/token drain 
& Router impersonation 
& Fake swap/LP add/remove sequences 
& DeFi \\

MEV-bait Phishing~\cite{morales2025mev} 
& MEV bots 
& Bot fund drain 
& False arbitrage signal 
& Malicious callback draining MEV bot 
& EVM-compatible \\

\midrule

% ----------------------- Ours -----------------------

\textbf{EIP-7702 Phishing (Ours)} 
& \textbf{Authorization layer} 
& \textbf{Execution hijack} 
& \textbf{Delegated-auth abuse} 
& \textbf{Malicious delegate installation via phished tuple} 
& \textbf{Ethereum (EIP-7702)} \\

\bottomrule

\end{tabular}%
}
\vspace{-0.8em}
\end{table*}

\smallskip
\noindent\textbf{The missing link.}
Despite its rapid uptake, the security consequences of EIP-7702 remain unexplored. Prior phishing research focuses on transaction-level deception, such as payload manipulation, approval hijacking, or malicious routers~\cite{slowmist2023permit,veritas_permit,hacken2023honeypot,yang2024stole}, which all require users to authorize \emph{harmful transactions}. In contrast, EIP-7702 enables a qualitatively different primitive: once a victim signs a single authorization tuple, every subsequent invocation of their account, regardless of origin, executes attacker-supplied bytecode. No prior work analyzes this semantic shift, determines whether it enables full-account compromise, or maps the conditions under which such compromise can be reliably triggered.

\smallskip
\noindent\textbf{An empirical study.}
Before establishing whether this mechanism is exploitable, it is essential to understand how EIP-7702 is being used in the wild: who controls the delegator contracts, how widely delegation is deployed, and whether observed endpoints already exhibit malicious patterns. To answer these questions, we conduct the first empirical measurement of 7702 authorization activity, covering over \textbf{150k} authorization and execution events across \textbf{26k+} addresses and \textbf{hundreds} of delegator contracts (Appendix~\ref{app:pilot_study}). Our results reveal burst-driven growth, extreme centralization, and heavy reuse of delegator contracts linked to known crime clusters, together with substantial real-world losses across ETH, ERC-20 tokens, and NFTs. These findings confirm that the delegation pathway is already being used at scale and that many real-world delegator endpoints function as malicious infrastructure.

\smallskip
\noindent\textbf{This work.}
Building on these observations, we present the first in-depth security analysis of EIP-7702 as a phishing vector. We show that a single malicious authorization tuple converts an EOA into a persistent proxy whose execution unconditionally resolves to attacker-controlled logic. Through controlled experiments, we demonstrate that full-account compromise can be triggered under three independent conditions: (\emph{i}) benign user-driven transactions, (\emph{ii}) attacker-driven external calls, and (\emph{iii}) ambient protocol callbacks. We further show that modern execution pipelines amplify these risks: ERC-4337 enables remote, repeatable activation through the \textsf{EntryPoint} during validation or execution, while chain-agnostic authorization (\texttt{chainId = 0}) enables replay-like compromise across independent networks.

\smallskip
\noindent\textbf{Contributions.}
This work makes the following contributions:
\begin{itemize}
    \item We present the first data-driven analysis of 7702 delegations, revealing a highly centralized, crime-dominated ecosystem underlying real-world adoption.
    \item We show that a single malicious authorization tuple enables total execution hijack of a victim EOA and validate three trigger conditions through end-to-end experiments.
    \item We uncover dangerous composability with ERC-4337 and show that chain-agnostic signatures enable replay-like compromise across independent networks. 
    \item We propose protocol-level and ecosystem-level defenses to constrain the delegation-based phishing attack surface.
\end{itemize}

% \qw{may get general inspirations from ndss25 \url{https://arxiv.org/pdf/2409.02386}}

% \noindent\textbf{Why 7702 phishing is harder for wallets to detect.}
% Phishing attacks enabled by EIP-7702 introduce challenges that exceed the detection capabilities of current wallet interfaces. Conventional wallets rely on transaction-level cues (e.g., destination addresses, transferred value, and modified allowances) to warn users about suspicious intent. EIP-7702 undermines this heuristic model by shifting the locus of risk from the transaction being executed to the persistent execution semantics of the account itself. 

%=================================================
\section{Technical Warmups}
%=================================================

\subsection{EIP-7702}
\noindent\textbf{Typed transactions for code delegation.}
EIP-7702~\cite{buterin2024eip} introduces a new EIP-2718 transaction type~\cite{eip2718} that allows an EOA to adopt contract-like behaviour without altering its address or deploying a new contract. The mechanism extends the typed-transaction framework with a dedicated \textit{set code transaction} (\texttt{0x04}) that embeds an \textit{authorization list} containing a set of signed delegation intents. When such a transaction is processed, the client interprets these intents as explicit instructions from the EOA’s owner to attach external logic to the account. This design provides a protocol-level pathway for EOAs to opt into enhanced programmability.

\smallskip
\noindent\textbf{Delegation through code substitution.}
The core idea of EIP-7702 is that an EOA may install a delegation marker (\texttt{0xef0100 || address})\footnote{The three-byte prefix \texttt{0xef0100} serves as a \emph{marker} indicating that the account code is not standard EVM bytecode but a delegation pointer. The first byte \texttt{0xef} corresponds to a deliberately banned opcode (per EIP-3541) to prevent accidental execution, while the following two bytes (\texttt{0x0100}) identify the format version and distinguish EIP-7702 delegations from other special code types.} in its code field that redirects execution to a target contract. Once the delegation is installed, any call to the EOA, whether originating from a user transaction or from another contract, executes the target code in the EOA’s context. This preserves the account’s storage, balance, and identity while enabling behaviour specified entirely by the delegated implementation. Importantly, the delegation does not modify code-introspection semantics: operations such as \texttt{EXTCODESIZE} and \texttt{EXTCODEHASH} continue to reflect the EOA’s own code field rather than the delegated implementation.

\smallskip
\noindent\textbf{Persistent and reversible behaviour upgrades.}
A distinguishing feature of EIP-7702 is its choice to make delegation persistent across transactions until explicitly revoked. Instead of restricting delegation to the lifetime of a single transaction, the proposal applies the change directly to the account’s code slot, ensuring that the EOA continues to behave as a smart wallet in subsequent interactions. This persistence offers a lightweight upgrade path: users can transition from an EOA to a programmable wallet implementation without deploying a contract, and can revert to a pure EOA state by authorizing a delegation to the zero address.

\subsection{Account Abstraction} 

The pursuit of Account Abstraction~\cite{wang2023account} aims to migrate users from  EOAs to programmable smart contract accounts. EIP-4337~\cite{erc4337} introduced a separate transaction flow (\texttt{UserOps}) and infrastructure (\texttt{Bundlers}) to achieve this, but it faces adoption challenges as most users remain on EOAs. EIP-3074~\cite{erc3074} attempted to bridge this by introducing opcodes (\texttt{AUTH}/\texttt{AUTHCALL}) allowing EOAs to delegate control to invoker contracts temporarily. However, this approach created a parallel ecosystem incompatible with the EIP-4337 architecture and required a hard fork for new opcodes.

\smallskip
\noindent\textbf{The EIP-7702 solution.} 
EIP-7702 aims to unify these approaches. Instead of introducing new opcodes or forcing a permanent migration (like EIP-5003), EIP-7702 introduces a new transaction type accepting a \texttt{contract\_code} field and a signature. During transaction execution, the EOA's code is temporarily set to the provided smart contract logic (enabling it to act as a contract wallet). 

This design offers advantages as follows.
\begin{packeditemize}
    \item \textit{Compatibility.} It allows EOAs to reuse existing EIP-4337 wallet implementations directly.
    \item \textit{Security.} Unlike permanent migration, the code assignment is transient per transaction, reducing the risk of accidental permanent loss of control, while maintaining the ability for future permanent upgrades if desired.
    \item \textit{No new opcodes.} It leverages existing \texttt{verify} and \texttt{execute} logic rather than adding opcode changes to the consensus layer.
\end{packeditemize}

\renewcommand{\arraystretch}{1.25}
\begin{table}[ht]
\centering
\caption{Comparison Among EIP-3074, EIP-4337, and EIP-7702}
\label{tab:eip-comparison}
\footnotesize
\begin{tabular}{
  c | c| c | c
}
\toprule
 \multicolumn{1}{c}{\textbf{Property}} & \multicolumn{1}{c}{\textbf{EIP-3074}} & \multicolumn{1}{c}{\textbf{EIP-4337}} & \multicolumn{1}{c}{\textbf{EIP-7702}} \\
\midrule
Execution Model     & \texttt{AUTHCALL} & Contract Wallet   & Delegated Code \\
Transaction Type    & Legacy + Opcode   & UserOp (0x2)   & Typed Tx (0x04) \\
EOA Compatible      & Yes               & No                & Yes \\
Delegation Type     & One-shot          & Full Control      & Persistent \\
State Change        & None              & Deploy Wallet     & Modify Code \\
Consensus Change    & Minor (Opcode)    & None              & Minor (Logic) \\
Security Concern    & Misuse of Call    & Bundler Trust     & Phishing Risk \\
\bottomrule
\end{tabular}
\end{table}

\subsection{Phishing Attack}
\noindent\textbf{Phishing attack.} 
Phishing attacks in blockchain systems typically exploit a user’s misunderstanding of what they are authorizing, rather than any flaw in the underlying cryptographic mechanisms~\cite{chen2024dissecting}. An adversary presents the victim with a transaction or signature request that appears benign but implicitly grants the attacker the ability to transfer assets or manipulate approvals. On Ethereum, this often takes the form of signing messages that encode ERC-20 approvals or permit-style authorizations, where the attacker abuses the legitimate semantics of the target contract. Because the victim intentionally produces a valid signature, these attacks bypass traditional security assumptions and produce on-chain transactions that appear indistinguishable from honest user activity. The reliability of phishing as an attack vector stems from two properties~\cite{chen2020phishing,li2022ttagn}: users cannot easily interpret the low-level effects of the operations they authorize, and wallets have limited ability to surface the full semantic meaning of signatures before they are submitted on chain.

\smallskip
\noindent\textbf{Phishing with 7702 (sketch).} 
EIP-7702 introduces a richer surface for phishing because delegation instructions are expressed as signed authorization tuples whose effects are not visible in the transaction fields that users typically examine. A malicious party can craft an authorization request that appears structurally harmless yet installs delegated code that carries arbitrary wallet logic. Once processed, the victim’s EOA behaves according to the attacker-controlled implementation, enabling downstream calls to drain assets or forward approvals without requiring further signatures from the victim. The attack does not rely on misleading contract calls or deceptive UI flows in existing dApps; instead, it leverages the fact that delegation is encoded in a compact pointer that reveals nothing about the behaviour of the underlying implementation. This disconnect between what the user sees (a short list of numerical fields) and what the protocol interprets (a persistent behaviour change at the account level) creates a fundamentally different form of phishing in which a single signed tuple can redefine how an account executes future transactions. 

The resulting attack surface consists of multiple variants that depend on how the unauthorized delegation is installed and subsequently activated, which we outline in \S\ref{sec:taxonomy}.

%=================================================
\section{Formalising Our Attacks}
\label{sec:formal-eip7702}
%=================================================

\subsection{Basics}

\noindent\textbf{Set-code transaction.} 
We model a set-code transaction as an RLP\footnote{\textbf{Recursive Length Prefix (RLP)} is Ethereum's canonical binary serialization format for encoding byte strings and nested lists; see \cite{wood2014ethereum_yellowpaper} for the formal specification.}-encoded tuple consistent with EIP-7702. For brevity, let
\[
\mathcal{T}_{\text{set}} = \mathrm{RLP}([\mathsf{tx\_fields},\ \mathsf{auth\_list},\ \mathsf{sig}]),
\]
where $\mathsf{tx\_fields}$ denotes the standard EIP-4844/EIP-1559~\cite{eip1559} fields (chain\_id, nonce, gas parameters, destination, value, data, access\_list, \ldots), $\mathsf{auth\_list}$ is the \emph{authorization list}, and $\mathsf{sig}=(y_{\mathrm{par}}, r, s)$ is the secp256k1 signature over the serialized payload. The authorization list is itself a vector of tuples:
\[
\mathsf{auth\_list} = \bigl[\,a_1, a_2, \ldots, a_k\,\bigr], \quad
a_i = \bigl(\mathit{cid}_i,\ \mathit{addr}_i,\ \nu_i,\ y_i,\ r_i,\ s_i\bigr).
\]

% \smallskip
% \noindent\textbf{Authorization tuple semantics.} 
% Each tuple $a=(\mathit{cid},\ \mathit{addr},\ \nu,\ y,\ r,\ s)$ encodes a signer-authorized intent. Define the signing message
% \[
% \mathsf{msg}( \mathit{cid}, \mathit{addr}, \nu ) \;=\; \mathcal{H}(\mathsf{MAGIC} \,\|\, \mathrm{RLP}([\mathit{cid},\ \mathit{addr},\ \nu])),
% \]
% and recover an authority address via
% \[
% \mathit{auth} \;=\; \mathrm{ecrecover}(\mathsf{msg}, y, r, s).
% \]
% Tuple $a$ is considered \emph{valid} in a transaction if the following checks hold (all arithmetic bounds per EIP-7702):
% \begin{enumerate}[label=(\alph*)]
%   \item $\mathit{cid}$ equals $0$ or the executing chain id;
%   \item $\nu < 2^{64}-1$ and matches $\mathrm{nonce}(\mathit{auth})$ at time of processing;
%   \item $r,s$ satisfy ECDSA canonicality constraints (EIP-2 style).
% \end{enumerate}

\smallskip
\noindent\textbf{Authorization tuple semantics.} 
Each authorization tuple captures a signed statement in which the user endorses a specific delegation target under well-defined protocol conditions. The fields appearing in the tuple reflect these commitments: the \textit{cid} value specifies the chain on which the authorization is intended to apply, the \textit{addr} field encodes the address of the delegated implementation, and the \textit{v} field represents the expected nonce of the authorizing account at the moment the client processes the tuple. These three components form the message to be signed, denoted as
\[
\mathsf{msg}(cid, addr, \nu)=\mathcal{H}(\mathsf{MAGIC}\,\|\,\mathrm{RLP}([cid,addr,\nu])),
\]
where the prefixed $\mathsf{MAGIC}$ byte distinguishes EIP-7702 authorizations from other signature domains. The signature components $(y,r,s)$ provide the parity bit and ECDSA signature values required for key recovery. The client thus derives the signer’s address via
\[
\mathit{auth}=\mathrm{ecrecover}(\mathsf{msg},y,r,s).
\]

A tuple is considered valid if its chain identifier matches the executing environment, if its nonce aligns with the current nonce of the recovered authority, and if the signature satisfies standard ECDSA canonicality checks. These conditions ensure that the tuple reflects a fresh, chain-specific authorization produced by the holder of the corresponding private key, and that it cannot be replayed across nonces or forged through malleable signatures.

When valid, processing $a$ causes the client to write a \emph{delegation indicator} into $\mathit{auth}$'s code slot:
\[
\mathrm{code}(\mathit{auth}) \leftarrow \underbrace{\texttt{0xef0100}}_{\text{marker}} \,\|\, \mathit{addr}.
\]

Processing occurs prior to the transaction’s execution phase but after sender nonce increment; delegation writes are \emph{not} reverted upon subsequent execution failure.

\subsection{Delegation State Transition}

\noindent\textbf{State model.} 
Let $\Sigma$ denote the global chain state. For any EOA $\mathit{e}$, let $\mathrm{code}_\Sigma(\mathit{e})$ denote its code bytes and $\mathrm{nonce}_\Sigma(\mathit{e})$ its nonce. Define the set of active delegations at block height $h$:
\[
\mathcal{D}_h \;=\; \{\, (\mathit{auth}, \mathit{target}) \mid \mathrm{code}_{\Sigma_h}(\mathit{auth}) = \texttt{0xef0100} \| \mathit{target}\,\}.
\]

Processing a set-code transaction $\mathcal{T}_{\text{set}}$ applies the transition
\[
\Sigma \xrightarrow{\mathcal{T}_{\text{set}}} \Sigma'
\]
such that for each valid tuple $a_i$ with recovered authority $\mathit{auth}_i$ and target $\mathit{addr}_i$:
\[
\mathrm{code}_{\Sigma'}(\mathit{auth}_i) \;=\; 
\begin{cases}
\texttt{0xef0100}\|\mathit{addr}_i, & \text{if } \mathit{addr}_i \neq \texttt{0x0},\\[2pt]
\text{empty\_code\_hash}, & \text{if } \mathit{addr}_i = \texttt{0x0}.
\end{cases}
\]
The global delegation set $\mathcal{D}$ is updated accordingly. When a tuple specifies $\mathit{addr}=\texttt{0x0}$, the client clears the account’s code by resetting its code hash to the \emph{empty\_code\_hash}, namely the canonical hash used in Ethereum to represent an account with no bytecode. This distinguishes genuine EOAs from delegated accounts that merely point to the zero address. Note that the transaction charges gas for all tuples irrespective of validity; refunds may apply per EIP-7702 rules.

\subsection{Attack Model}

We consider three classes of entities relevant to EIP-7702 phishing. The \emph{victim} EOA $V$ represents an honest user operating through standard wallet software. The \emph{attacker} $A$ controls off-chain infrastructure and one or more on-chain addresses intended to receive stolen assets. Finally, auxiliary \emph{infrastructure} $R$ consists of wallet clients, browser extensions, relayers, or bundlers that forward transactions on behalf of users. These components may behave honestly or negligently but do not collude with victims.

\smallskip
\noindent\textbf{Adversary capabilities.}  
The attacker’s ability to induce unauthorized delegation follows from realistic vectors observed in existing phishing attacks. The adversary can craft deceptive interfaces or prompts that elicit a signature from $V$ over the authorization message $\mathsf{msg}(cid, addr, \nu)$, thereby obtaining a valid tuple $(y, r, s)$ without revealing its behavioural implications. In stronger scenarios, the attacker may have partial or full access to $V$'s private key, enabling the creation of arbitrary authorization tuples without user involvement. The adversary can deploy contracts that serve as delegated implementations, broadcast set-code transactions (\texttt{AuthTx}), and subsequently issue arbitrarily crafted follow-up transactions that exploit the installed delegation. The attacker may monitor the mempool and automate sweeps to ensure rapid inclusion, but cannot violate consensus rules or break signature schemes.

\smallskip
\noindent\textbf{Attacker objective.}  
The attacker aims to modify the victim’s account semantics rather than any single transaction outcome. The central objective is to cause a \textbf{persistent delegation write} that installs an attacker-controlled implementation $T$ as the execution target for $V$. Once the delegation indicator is in place, the adversary gains the ability to shape how $V$ behaves in subsequent execution contexts: any transaction that addresses $V$, whether initiated by the victim, a dApp, or an external contract, is interpreted through the logic contained in $T$. This shift grants the attacker a durable and unilateral capability to issue operations on behalf of $V$ without further signatures, enabling arbitrary contract interactions to be carried out under $V$’s authority whenever a suitable trigger occurs.

\smallskip
\noindent\textbf{Attack lifecycle.}
An EIP-7702 phishing attack can be modelled as a short sequence of on-chain events:
\[
\begin{aligned}
\texttt{AuthTx} & \xrightarrow{\text{process}} \mathsf{write\_delegation}(\mathit{auth}=V,\ \mathit{target}=T) \\
                & \xrightarrow{\text{trigger}} \texttt{ExploitTxs} \\
                & \xrightarrow{\text{settle}} \texttt{SweepTxs}
\end{aligned}
\]
subject to the precondition $T \neq \texttt{0x0}$. Here:

\begin{packeditemize}
  \item \texttt{AuthTx} is a set-code transaction that contains an authorization tuple whose recovered signer (via \texttt{ecrecover}) equals the victim $V$. This tuple may originate from a legitimate signature by $V$, or from a forged one if $V$'s private key is compromised.
  
  \item $\mathsf{write\_delegation}$ denotes the client-side update that writes the delegation indicator $\texttt{0xef0100} \,\|\, T$ into $\mathrm{code}(V)$. 
  
  \item \texttt{ExploitTxs} refers to subsequent transactions that, when executed against $V$, are routed to $T$ under delegation semantics. These may be initiated by $V$ or by third parties, typically performing asset transfers (e.g., \texttt{transferFrom}, ETH sends).
  
  \item \texttt{SweepTxs} finalize the exploit by consolidating assets at attacker-controlled addresses, possibly including swaps or cross-chain bridges to obscure provenance.
\end{packeditemize}

\smallskip
\noindent\textbf{Condition for successful drain.} 
Let $\mathrm{bal}_\Sigma(X)$ denote the on-chain ETH/ERC-20 balances associated with account $X$ at state $\Sigma$. A successful drain by target $T$ exists between blocks $h$ and $h'$ if there exists a sequence of transactions $\{ \tau_t \}_{t\in[h,h']}$ such that:
\[
\mathrm{bal}_{\Sigma_{h'}}(V) < \mathrm{bal}_{\Sigma_{h}}(V)
\quad\text{and}\quad
\sum_{t} \mathrm{value\_to\_A}(\tau_t) \ge \Delta,
\]
where $\Delta$ represents a threshold for economically meaningful loss in our measurement. For some appreciable $\Delta>0$ and where the execution of at least one $\tau_t$ dereferences $\mathrm{code}(V)$ via the delegation indicator and invokes transfer logic directing funds to $A$.

%================================================= 
\section{Attack Methodology}
\label{sec-methodology}
%================================================= 
% This section describes how the conceptual model of EIP-7702 phishing is instantiated as an operational attack pipeline. We begin by organizing the space of EIP-7702 phishing behaviours into a taxonomy that captures the structural patterns attackers exploit. We then detail the construction of malicious delegated implementations, the generation of authorization tuples, and the formation of set-code transactions that install these tuples on-chain. Finally, we explain how delegated execution is triggered and how draining strategies are orchestrated in a controlled environment.

\subsection{Phishing Taxonomy under EIP-7702.} 
\label{sec:taxonomy}
Phishing under EIP-7702 represents a family of attacks that exploit the semantic gap between the compact authorization tuples a user signs and the persistent behavioural changes these tuples induce on the account. 
\begin{packeditemize}
    \item The \underline{first} class of attacks targets the delegation pointer itself: users are deceived into authorizing an innocuous address that in fact hosts attacker-controlled wallet logic. Because the delegated implementation is opaque at the time of signing, victims have no practical way to evaluate the behaviour they are installing. 
    
    \item A \underline{second} class arises from misuse of persistence. Once delegation is written to the account’s code slot, subsequent interactions, many of which appear unrelated to the initial authorization, are automatically routed through the attacker’s code. This creates a temporal separation between the phishing moment and the asset-draining moment, enabling attackers to operate opportunistically long after the victim has forgotten the initial action. 
    
    \item A \underline{third} class concerns cross-context execution. Delegated code is triggered not only by transactions sent by the victim but also by external calls from other contracts, which allows an attacker to craft on-chain environments that indirectly activate the malicious implementation. This expands the phishing surface beyond the user interface and into protocol interactions that the victim never observes directly. 
\end{packeditemize}

Across those, the unifying factor is that EIP-7702 grants a single signed tuple the ability to redefine an account’s execution semantics, allowing attackers to shift from manipulating transaction parameters to manipulating the account’s underlying behaviour.

\begin{figure}[t]
  \centering
  \includegraphics[width=0.99\linewidth]{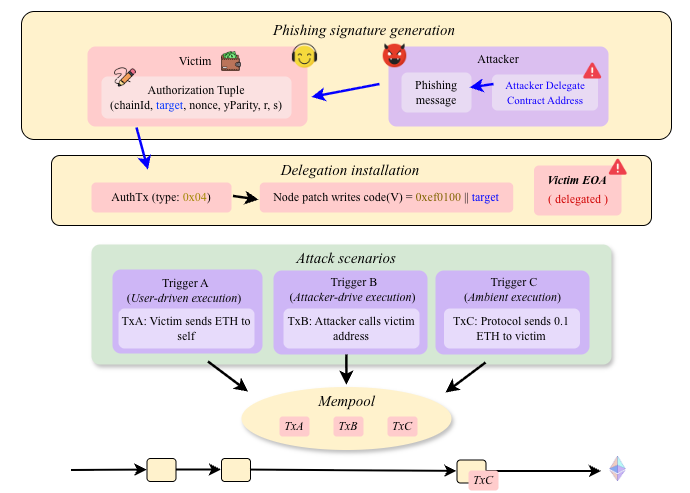}
  \caption{\textcolor{teal}{EIP-7702 phishing attack.} The victim signs an authorization payload that installs a delegation pointing to an attacker-controlled contract. A later call resolves delegation and runs code in the victim’s context, allowing the attacker to issue approvals or transfer assets.}
  \label{fig:workflow_eip7702}
  \vspace{-0.1in}
\end{figure}

\subsection{Constructing Malicious Delegations}
\noindent\textbf{Designing the execution entry point.}  
A malicious EIP-7702 delegation begins with the construction of a contract whose runtime code activates immediately when invoked through delegated execution. Since the victim’s address becomes \texttt{msg.sender}, the implementation must treat every entry as implicitly authorized and avoid dependence on structured calldata. Practically, this is achieved by placing the draining logic inside the fallback function, ensuring it executes under any invocation path, including calls originating from unrelated dApps or protocol-level interactions. The entry routine typically retrieves token balances using \texttt{balanceOf}, issues \texttt{transfer} or \texttt{transferFrom} calls to attacker-controlled sinks, and forwards ETH using low-level calls that propagate all available gas. This design guarantees that the malicious behaviour triggers even under minimal calldata or when invoked by accident through routine user activity.

\smallskip
\noindent\textbf{Implementing draining logic.}  
Delegated execution grants the attacker access to the victim’s storage and approval relationships, enabling contract logic that interacts with ERC-20, ERC-721, and DeFi protocols as if the victim had initiated the calls directly. Malicious implementations therefore incorporate routines for sweeping multiple asset types, including token transfers, NFT transfers, and swaps that convert illiquid assets into ETH before extraction. To improve robustness, the draining logic avoids assumptions about token behaviour, handles failing transfers through low-level calls, and may include multi-hop pathways via attacker-controlled routers. Because delegated execution inherits the victim’s authority, no impersonation checks are required, making these operations indistinguishable from legitimate wallet interactions at the EVM level.

\smallskip
\noindent\textbf{Deploying reusable malicious templates.}  
The attacker must deploy the delegated implementation as a regular contract with non-empty bytecode, avoiding precompile addresses to satisfy client validation rules. Beyond these constraints, the design space is flexible. Attackers often favour compact bytecode to minimize deployment cost and avoid attracting attention during the phishing flow. A common strategy is to pre-deploy a library of malicious templates, each optimized for a specific extraction strategy such as ERC-20 draining, NFT transfers, or multi-step DeFi swaps. Reusing these templates allows the attacker to embed only a static address in authorization tuples, reducing on-chain artifacts during the actual phishing attempt and simplifying large-scale campaigns. Once deployed, the contract address serves as the persistent execution target installed on the victim’s account via the authorization tuple.

\subsection{Crafting Authorization Tuples}
\noindent\textbf{Constructing valid signatures.}  
To successfully induce a malicious delegation, the attacker needs to construct authorization tuples that pass all client-side validation while appearing benign to the victim’s wallet interface. This requires producing a signature over the delegation message using either a \textit{phished signature} or a \textit{compromised private key}, ensuring the tuple fields (chain identifier, target address, and expected nonce) match the current on-chain state. A practical phishing flow therefore begins with presenting the victim with an approval request disguised as a routine session message or application login, prompting them to sign data that reveals neither operational semantics nor code-level intent. When operating without user interaction, compromised-key scenarios allow attackers to mass-produce tuples programmatically, synchronizing tuple nonces with live chain state.

\smallskip
\noindent\textbf{Optimize tuple parameters.} Beyond validity, attackers optimize tuple parameters to avoid detection and maximize stealth. The target field embeds the address of the attacker’s delegated implementation, often selected to resemble legitimate module or wallet contracts so that the victim sees only an unremarkable address rather than executable logic. Attackers frequently reuse the same target address across many victims, enabling scalable automation and reducing on-chain noise. Chain identifiers may be set to zero to widen cross-chain applicability, while tuples are arranged so that their ordering respects nonce increments during preprocessing.

\subsection{Embedding Tuples into AuthTx}
The core task for the attacker is to embed the collected authorization tuple into a type-4 set code transaction in a way that ensures the client will process it and write the delegation indicator into the victim’s account. The attacker serializes the tuple into the authorization list of the transaction and constructs the payload so that it satisfies all syntactic requirements of a valid set code transaction. This involves populating chain identifiers, gas fields, and outer transaction parameters so that the transaction is accepted by nodes and appears structurally routine when displayed by the wallet.

A key element of the attack is that clients process the authorization list before execution, and the resulting delegation write is not reverted even if the later execution fails. The attacker exploits this by embedding the malicious tuple in a transaction that does not need to succeed semantically; the outer call may be a dummy call, a benign interaction, or even structured to revert. As long as the preprocessing step runs, the attacker’s delegation is installed. 

Wallets typically render only the outer transaction fields and do not display the authorization list. The attacker takes advantage of this by shaping the \texttt{AuthTx} so that the victim sees an operation consistent with familiar UX patterns, such as a harmless approval or a verification message. The authorization tuple remains hidden inside the payload, and embedding it in a syntactically valid structure allows the attacker to deliver the malicious delegation without triggering wallet-level warnings.

\subsection{Triggering Delegated Execution}
Once the malicious delegation has been written into the victim’s code slot, any call that reaches the victim’s address activates the attacker-controlled implementation. This property yields three distinct activation pathways that correspond to the experimental scenarios evaluated later in this work(\S\ref{sec-experiments}).

\smallskip
\noindent\textbf{\ding{172} User-driven activation.}  
The most natural trigger occurs when the victim initiates a routine transaction through a dApp or wallet interface. Because delegated execution replaces the EOA’s behaviour at the protocol level, any outbound call that targets the victim’s own address, or any contract interaction in which the victim is the designated sender, immediately dispatches into the malicious implementation. This activation path simulates realistic user activity and demonstrates that no additional attacker interaction is required once the delegation is installed.

\smallskip
\noindent\textbf{\ding{173} Attacker-driven activation.}
Delegated execution also applies to external invocations. An attacker can therefore trigger the malicious contract directly by sending a call to the victim’s address, even if the victim is offline. This pathway shows that the attacker can unilaterally initiate the draining phase after the phishing step completes, exploiting the fact that the victim’s account now behaves as a contract whose code is supplied by the adversary.

\smallskip
\noindent\textbf{\ding{174} Ambient execution.}
A third pathway arises from protocol-level interactions that involve the victim as part of background system activity. Examples include token callbacks, settlement routines, or auxiliary contract calls issued by automated services. Any such call that targets the victim implicitly resolves to the delegated implementation. This pathway demonstrates that the attacker can rely on ordinary on-chain activity to trigger the malicious fallback, allowing the draining routine to activate without either the victim or the attacker issuing an explicit transaction.

Across these activation paths, the critical observation is that EIP-7702 transforms a phishing event into a \emph{persistent} execution modification. Once installed, the malicious delegation will be invoked by whichever trigger occurs first, enabling flexible and opportunistic extraction strategies.

\begin{algorithm}[t]
\caption{EIP-7702 Phishing Attack}
\label{alg:eip7702-attack-ultra}
\smallskip
\begin{algorithmic}[1]
\Require Victim EOA $V$; attacker implementation $T$; chain state $\Sigma$

\Statex \textcolor{magenta}{\textbf{Delegate preparation:}}
\State Deploy $T$ with fallback: enumerate asset balances; invoke ERC-20 \texttt{transfer/transferFrom}; ERC-721 \texttt{transferFrom}; forward ETH via \texttt{call(gas())}; perform optional AMM swaps.

\Statex \textcolor{magenta}{\textbf{Tuple construction:}}
\State $n \gets \mathrm{nonce}(V)$ from $\Sigma$
\State $\mathrm{msg} \gets \mathrm{keccak}( \mathrm{MAGIC} \,\|\, \mathrm{RLP}[cid, T, n] )$
\State Obtain $(y,r,s)$ via phishing signature or compromised key
\State $a \gets (cid, T, n, y, r, s)$

\Statex \textcolor{magenta}{\textbf{AuthTx assembly:}}
\State Build type-4 transaction $\mathcal{T}_{set}$:
\Statex \hspace{1.5em}-- embed $a$ into \texttt{authorization\_list}
\Statex \hspace{1.5em}-- fill outer fields: $(\mathrm{gas}, \mathrm{to}, \mathrm{data}, \mathrm{access\_list})$
\State Broadcast $\mathcal{T}_{set}$

\Statex \textcolor{magenta}{\textbf{Delegation write:}}
\State Client verifies tuple $a$
\State Write delegation:
\[
\mathrm{code}(V) \gets \texttt{0xef0100} \,\|\, T
\]

\Statex \textcolor{magenta}{\textbf{Execution trigger:}}
\State Any call to $V$ (user tx, external contract, attacker ping):
\[
\mathrm{dispatch}(V) \rightarrow T
\]

\Statex \textcolor{magenta}{\textbf{Drain routine:}}
\State $T$ executes under $V$'s authority:
\Statex \hspace{1.5em}-- fetch ERC-20 balanceOf; execute token drains  
\Statex \hspace{1.5em}-- move NFTs via \texttt{transferFrom}  
\Statex \hspace{1.5em}-- swap assets → ETH  
\Statex \hspace{1.5em}-- forward value to attacker addresses

\end{algorithmic}
\end{algorithm}

% \subsection{Limitations and ethical notes}
% The attack requires an explicit signature or consent from the victim; it does not circumvent cryptographic protections. Widespread improvements to wallet UX and explicit per-tuple consent could substantially reduce practical feasibility. All experiments described in this work should be performed on testnets or controlled private forks; responsible disclosure procedures should be followed for any live incidents.

%================================================= 
\section{Experiments}
\label{sec-experiments}
%================================================= 
We instantiate the full attack pipeline in a controlled local environment and evaluate the practicality of our attack.

\subsection{Experimental Setup}
\label{sec:exp-setup}

All experiments are conducted on a macOS system (Node.js v20.18.2, Hardhat v2.27.0) with Solidity~0.8.24 and optimizer enabled. Hardhat serves as the execution backend, providing a local Ethereum network with full tracing support and deterministic account generation.

\smallskip
\noindent\textbf{Network configuration.}  
We run a Hardhat local network configured with chain identifier 1337, Cancun hardfork rules, and unrestricted contract size to accommodate research artifacts. The genesis state is empty, and all experiments begin at block height~0. Hardhat’s deterministic account set is used to fix the victim and attacker addresses, each pre-funded with 10,000~ETH to ensure reproducibility. Logging and transaction tracing are enabled throughout.

\smallskip
\noindent\textbf{EIP-7702 enablement.}  
Since Hardhat does not natively implement transaction type~0x04, we extend its execution environment with a preprocessing layer that emulates the EIP-7702 state transition rules. The patch modules (\texttt{patchEVM.js} and \texttt{patchHardhatNode.js}) intercept incoming type-4 transactions, extract the authorization\_list, validate tuple structure, compute the authority address, and write the corresponding delegation indicator (\texttt{0xef0100}~\textbar\textbar~\textit{target}) into the authority account’s code before execution begins. The patch integrates with Hardhat’s environment extension mechanism and is automatically applied on node startup.

\smallskip
\noindent\textbf{Contracts and deployment.}  
The attacker’s implementation is deployed as a compact malicious fallback contract, which executes draining logic upon any delegated call  (i.e., \texttt{MaliciousDelegate.sol}). A mock ERC-20 contract (\texttt{MockERC20.sol}) is deployed to emulate fungible assets, and a dummy protocol contract is provided to test indirect (ambient) triggers. All deployments are deterministic: contract addresses, bytecode hashes, and deployment transactions are recorded for reproduction.

\smallskip
\noindent\textbf{Attack tooling.}  
We implement a suite of scripts to orchestrate the attack pipeline. 
\begin{packeditemize}
    \item \texttt{generateTuple.js} constructs valid authorization tuples by computing the EIP-7702 signing message and producing the corresponding secp256k1 signature. 
    \item \texttt{sendAuthTx.js} builds and broadcasts type-4 transactions embedding the tuple, installing the malicious delegation on the victim account.  
    \item \texttt{trigger.js} executes all activation scenarios (user-driven, attacker-driven, and ambient triggers) and instruments each transaction using \texttt{debug\_traceTransaction}. 
    \item \texttt{runAndCollect.js} aggregates traces, state diffs, balance snapshots, and logs into structured JSON artifacts.
\end{packeditemize}

% \paragraph{Data collection and reproducibility.}  
% For every experiment, we collect execution traces, storage diffs, authorization tuple encodings, delegation installation evidence, and balance snapshots. All output is written to the \texttt{experiments/} directory, including a consolidated report summarizing the complete experiment. The environment uses fixed private keys, fixed chain identifiers, deterministic nonces, and deterministic deployments to ensure exact reproducibility across runs.

\subsection{Experimental Execution}

\noindent\textbf{Phase 1: Delegation installation.}
The attacker installs a malicious delegation on the victim account by embedding an authorization tuple into a type~\texttt{0x04} set-code transaction. The tuple is generated offline and encoded in RLP form; the patched Hardhat node interprets this as an EIP-7702 authorization and writes the delegation indicator into the victim's code slot.

The attacker invokes the following script to build and submit the set-code transaction:
\vspace{-0.12in}
\begin{lstlisting}[language=javascript,caption={Broadcasting the EIP-7702 delegation AuthTx}]
const txHash = await provider.send("eth_sendTransaction", [{
  from: victimAddress,
  type: "0x04",
  gas: "0x500000",
  maxFeePerGas: "0x2540be400",
  maxPriorityFeePerGas: "0x2540be400",
  authorization_list: [
    // RLP-decoded tuple:
    // [chainId, target, nonce, yParity, r, s]
    [
      "0x539", 
      "0x5fbdb2315678afecb367f032d9...4180aa3",
      "0x00",
      "0x00",
      "0x8b1629fcc38442e2899e3c...2c5c5ef20",
      "0x1ebf532ec7c60bc6d10e8...8d18fbb346"
    ]
  ],
  to: victimAddress,
  data: "0x"
}]);

\end{lstlisting}

\noindent Upon execution, the patched node extracts the authorization list, validates the tuple format, verifies the signature using \texttt{ecrecover}, and writes the delegation indicator:
\[
\texttt{code}(V) \leftarrow \texttt{0xef0100} \Vert \textcolor{teal}{adr.0aa3}.
\]

The node-side logs confirm that the preprocessing occurred before any transaction execution:
\vspace{-0.12in}
\begin{lstlisting}[language=bash,caption={Node logs confirming correct EIP-7702 preprocessing}]
[7702] detected type-0x04 transaction
[7702] extracted authorization_list (1 tuple)
[7702] tuple signature verified: signer = 
       0xf39fd6e51aad88f6f4c...9cfffb92266
[7702] writing delegation code:
       0xef01005fbdb2315678a...642f64180aa3
[7702] delegation write persisted (revert-safe)
\end{lstlisting}
\vspace{-0.12in}
% We verify that the victim’s account code transitioned from empty to the expected delegated pointer:
% \begin{lstlisting}[language=bash,caption={Delegation indicator successfully written to victim account}]
% $ cast code 0xf39fd6e51aad88f6f4ce6ab8827279cfffb92266
% 0xef01005fbdb2315678afecb367f032d93f642f64180aa3
% \end{lstlisting}

\noindent\textbf{Phase 2: Trigger A Attack (User-driven execution).}
Once the delegation indicator has been installed, we first evaluate whether an ordinary user-initiated transaction is sufficient to activate the malicious delegate. This scenario models the most common real-world trigger condition: a benign wallet action that unintentionally dispatches execution into attacker-controlled logic. The victim sends ETH to their own address, which is a routine pattern in many applications (e.g., force-sending gas, resetting a stuck nonce). Since any call targeting the victim's address now resolves to the delegated implementation, this transaction reliably enters the fallback function of the malicious contract.

This scenario is executed \emph{first} in the experiment sequence ($\textbf{A}\rightarrow B\rightarrow C$). Thus, the victim begins with the full initial balance\footnote{We denote ETH and ERC-20 test tokens with \ethsym,\,\ercsym~ for simpliciy.}: 

• \textbf{ETH:} \texttt{10000\,\ethsym (9999.997701772909192554\,\ethsym after gas fees)}  

• \textbf{Tokens:} \texttt{2000.0\,\ercsym}  

These values define the largest drainage window among all cases.

\begin{figure}[!htbp]
    \centering

    %--------------------------------------------------
    \subfigure[\textbf{Scenario A: User-driven trigger.}
    The victim voluntarily sends a transaction to itself, activating the delegated wallet code installed via EIP-7702. 
    The execution drains \textbf{9{,}999.9911 ETH} and \textbf{2{,}000 tokens} in a single run, leaving only the gas-dust remainder 
    (\texttt{eth\_after = 0.006564551451274734}). 
    This corresponds to transaction \textcolor{teal}{txid.b64f}.]{
        \label{fig:scenario-a}
        \includegraphics[width=\linewidth]{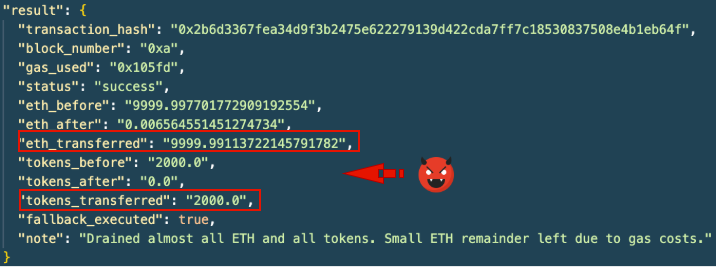}
    }

    \vspace{-0.08in}

    %--------------------------------------------------
    \subfigure[\textbf{Scenario B: Attacker-driven trigger.}
    After Scenario A, the victim’s balance has been reduced to dust. 
    The attacker then directly calls the victim’s address, which again executes the attacker’s delegate contract.
    This drains the remaining \textbf{0.006564551451274734 ETH}, bringing the balance fully to zero. 
    This corresponds to transaction \textcolor{teal}{txid.bb1f}.]{
        \label{fig:scenario-b}
        \includegraphics[width=\linewidth]{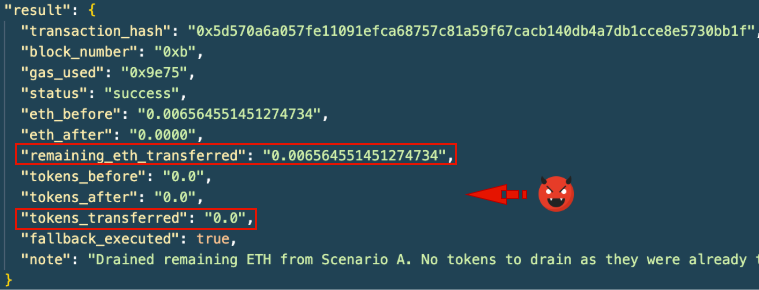}
    }

    \vspace{-0.08in}

    %--------------------------------------------------
    \subfigure[\textbf{Scenario C: Ambient trigger.}
    A benign protocol contract transfers \textbf{0.1 ETH} to the victim as part of a simulated workflow. 
    The transfer immediately activates the victim’s delegated execution and drains the full amount. 
    This corresponds to transaction \textcolor{teal}{txid.b0c1}, demonstrating that \textit{any} call path, not only active user actions, can activate the malicious logic.]{
        \label{fig:scenario-c}
        \includegraphics[width=\linewidth]{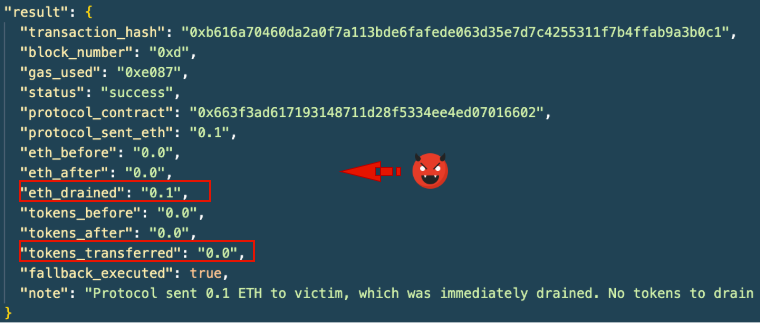}
    }

    \vspace{-0.1in}
    \caption{Execution traces produced during the EIP-7702 phishing attack across three activation pathways. }
    \vspace{-0.15in}
\end{figure}

\smallskip
\noindent\textit{\underline{Trigger transaction.}}
The victim initiates the following self-call, which Hardhat interprets as a standard \texttt{eth\_sendTransaction} request:
\vspace{-0.12in}
\begin{lstlisting}[language=javascript,caption={User-triggered ETH transfer to delegated account}]
const victimTx = {
    from: "0xf39fd6e...cfffb92266",   // adr.2266
    to:   "0xf39fd6e...cfffb92266",   // adr.2266
    value: ethers.utils.parseEther("0.1").toHexString(),
    gas: "0x500000"
};
// This transaction corresponds to txid.b64f
const txHash = await provider.send("eth_sendTransaction", [victimTx]);
\end{lstlisting}
\vspace{-0.12in}
% \smallskip
\noindent\textit{Trigger A result.}
Hardhat's execution trace confirms that the call no longer executes the EOA’s default empty code but immediately enters the fallback routine of the malicious delegate (\textcolor{teal}{adr.0aa3}). The trace shows internal calls for both ETH transfer and ERC-20 sweeping. We extracted the relevant outcomes:

\begin{itemize}
    \item The fallback routine executed successfully (\textit{verified via} \textit{debug\_traceTransaction}).
    \item All ETH in the victim account, except a minimal ETH dust remainder due to gas usage, was transferred to the sink address (\textcolor{teal}{adr.79c8}).
    \item All ERC-20 token balances (2{,}000\,\ercsym) were transferred via the delegate’s ERC-20 loop.
\end{itemize}
% \vspace{-0.12in}
\smallskip
\noindent\textit{Observed balance changes (Fig.~\ref{fig:scenario-a}).}
The experiment framework recorded pre/post state snapshots:
\vspace{-0.12in}
\begin{lstlisting}[caption={Balance differences for user-driven trigger}]
ETH transferred:     9999.99113722145791782
ERC-20 transferred:  2000.0
fallback_executed:   true
\end{lstlisting}
\vspace{-0.12in}
\noindent These results confirm that the user-driven trigger reliably activates delegated execution and immediately results in full asset drainage, demonstrating that no attacker-initiated follow-up is required once the victim performs any interaction that routes execution through their own address.

\smallskip
\noindent\textbf{Phase 3: Trigger B Attack (Attacker-driven execution, after A).}
After the full-drain event in Scenario~A, we next evaluate whether delegated execution remains active when the attacker, rather than the victim, initiates a transaction toward the compromised EOA. This scenario models an attacker probing the victim address directly, without requiring any further user interaction. Importantly, because the delegation indicator persists in the account’s code slot, any externally-originating call \emph{continues} to route execution into the malicious delegate.

This scenario is executed \emph{second} in the sequence ($A\rightarrow \textbf{B}\rightarrow C$). Consequently, the victim begins Scenario~B with only the residual ETH left after the first drain and no remaining tokens:

• \textbf{ETH:} \texttt{0.006564551451274734\,\ethsym}  

• \textbf{Tokens:} \texttt{0.0\,\ercsym}  

These values represent the minimal post-Scenario~A balance state.

\smallskip
\noindent\textit{\underline{Trigger transaction.}}
The attacker issues a direct call to the victim’s address using an empty calldata payload. Although this call would normally do nothing on a standard EOA, the presence of the EIP-7702 delegation indicator causes Hardhat’s patched EVM to resolve the execution target to the delegate contract:
\vspace{-0.12in}
\begin{lstlisting}[language=javascript,caption={Attacker external call to delegated victim (txid.\textcolor{teal}{bb1f}).}]
const attackerTx = {
    from: "0x7099797...d17dc79c8",  // adr.79c8
    to:   "0xf39fd6e...cfffb92266", // adr.2266
    data: "0x",
    gas:  "0x500000"
};
// This transaction corresponds to txid.bb1f
const txHash = await provider.send("eth_sendTransaction", [attackerTx]);
\end{lstlisting}
\vspace{-0.12in}
Because the delegated implementation places all attack logic in its \texttt{fallback()} function, the absence of calldata is immaterial, that is, any inbound call is sufficient to trigger the drainage routine.

\smallskip
\noindent\textit{Trigger B result.}
Hardhat's execution trace again confirms a full delegated dispatch into the malicious fallback function (\textcolor{teal}{adr.0aa3}). Since Scenario~A had already transferred all ERC-20 tokens and almost all ETH, the only remaining operation is the extraction of the last fraction of ETH dust. The fallback routine performs a successful low-level \texttt{call()} to the sink address (\textcolor{teal}{adr.79c8}) with the victim's remaining balance.

\begin{itemize}
    \item Delegated fallback execution was triggered correctly (\textit{verified via} \texttt{debug\_traceTransaction}).
    \item The remaining ETH balance (\texttt{0.006564551451274734\,\ethsym}) was fully transferred to the sink.
    \item No ERC-20 transfers were executed since the token balance was already zero.
\end{itemize}

\smallskip
\noindent\textit{Observed balance changes (Fig.~\ref{fig:scenario-b}).}
The experiment infrastructure recorded the pre/post balance state as follows:
\vspace{-0.12in}
\begin{lstlisting}[caption={Balance differences for attacker-driven trigger}]
ETH transferred:     0.006564551451274734
ERC-20 transferred:  0.0
fallback_executed:   true
\end{lstlisting}
\vspace{-0.12in}
\noindent This confirms that delegated execution remains fully active after Scenario~A. The attacker can continue to drain any amount of ETH (including newly received ether) without any additional signatures or wallet interaction from the victim. Scenario~B demonstrates that EIP-7702 phishing confers \emph{persistent and attacker-triggerable} control over the compromised account.

\smallskip
\noindent\textbf{Phase 4: Trigger C Attack (Ambient execution after A and B).}
The final scenario evaluates the most subtle activation pathway: an \emph{ambient call} made by a third-party contract. This models real-world situations where protocol components (routers, settlement modules, reward distributors, NFT marketplaces, or vault controllers) call back into a user’s address as part of routine system behavior. Under EIP-7702 delegation, such calls are indistinguishable from the victim’s own execution context and therefore silently enter the attacker-controlled implementation.

This scenario is executed \emph{last} in the sequence ($A \rightarrow B \rightarrow \textbf{C}$). At this stage, the victim’s account has already been fully drained by Scenarios A and B, leaving:

• \textbf{ETH:} \texttt{0.0\,\ethsym}  

• \textbf{Tokens:} \texttt{0.0\,\ercsym}  

Thus, the protocol-initiated call becomes the only new value injection into the victim’s delegated account. The DummyProtocol contract (\textcolor{teal}{adr.6602}) sends \texttt{0.1 ETH} to the victim as part of its call routine, providing a controlled environment to observe whether the attacker’s fallback logic still captures externally provided funds.

\smallskip
\noindent\textit{\underline{Trigger transaction.}}
The DummyProtocol contract exposes a simple entry point, \texttt{callTarget(address)}, which forwards ETH and issues a low-level call to the victim's address. The attacker or an unaware protocol invokes it as follows:
\vspace{-0.12in}
\begin{lstlisting}[language=javascript,caption={Ambient trigger via protocol call (txid.\textcolor{teal}{b0c1}).}]
const protocolCallTx = {
    from: callerAddress,
    to:   "0x663f3ad6...d07016602", // adr.6602
    data: dummyProtocol.interface.encodeFunctionData(
              "callTarget",
              ["0xf39f...b92266"] // adr.2266
          ),
    value: ethers.utils.parseEther("0.1").toHexString(),
    gas:   "0x500000"
};
// This transaction corresponds to txid.b0c1
const txHash = await provider.send("eth_sendTransaction", [protocolCallTx]);
\end{lstlisting}
\vspace{-0.12in}
The ambient call causes the protocol contract to attempt a low-level \texttt{target.call\{\}} invocation, which under 7702 delegation routes directly into the malicious delegate’s \texttt{fallback()} routine.

\smallskip
\noindent\textit{Trigger C result.}
Execution traces (\texttt{debug\_traceTransaction}) confirm that:

\begin{itemize}
    \item The protocol contract successfully delivered \texttt{0.1 ETH} to the victim's address.
    \item The delegated fallback executed immediately without any interaction from the victim.
    \item All \texttt{0.1 ETH} provided by the protocol contract was transferred to the attacker sink (\textcolor{teal}{adr.79c8}).
    \item No ERC-20 transfers occurred, as all tokens were already drained in Scenario A.
\end{itemize}

This scenario conclusively demonstrates that any arbitrary contract call (e.g., benign, unaware, or automated) can unintentionally activate the attacker’s code.

\smallskip
\noindent\textit{Observed balance changes (Fig.~\ref{fig:scenario-c}).}
The experiment’s data collection framework recorded:
\vspace{-0.12in}
\begin{lstlisting}[caption={Balance differences for ambient trigger.}]
protocol_sent_eth:   0.1
eth_drained:         0.1
tokens_transferred:  0.0
fallback_executed:   true
\end{lstlisting}
\vspace{-0.12in}
\noindent The fallback logic remains active even after all funds have been drained in earlier phases, meaning that \emph{any future ETH or tokens routed to the victim will be captured immediately and irreversibly}. This property makes ambient triggers particularly dangerous, as many on-chain workflows generate such callbacks without explicit user involvement.

%================================================= 
\section{Extended Attack Surfaces}
%================================================= 
\subsection{\textcolor{magenta}{(a)} EIP-7702 + ERC-4337 Composite Attack}
EIP-7702 replaces the static semantics of EOAs by allowing them to behave as dynamically delegated contract wallets, whereas ERC-4337 introduces an alternative transaction flow where \textsf{UserOperation} objects, rather than EOA-signed transactions, define the primary unit of execution. When combined, these two designs yield a compositional attack surface in which a compromised EIP-7702 delegation can be remotely and repeatedly activated through ERC-4337’s programmatic execution pipeline, without requiring any participation from the victim.

\smallskip
\noindent\textbf{Trigger surface expansion via \textsf{EntryPoint}.}
Under EIP-7702 alone, the attacker must rely on explicit calls to the victim’s address (either user-driven or ambient). ERC-4337 removes this constraint entirely. A malicious \textsf{UserOperation}, bundled by an attacker-controlled or economically-incentivized bundler, can force the \textsf{EntryPoint} to issue internal calls to the victim address during:

\begin{enumerate}[label=(\alph*)]
    \item \textsf{validateUserOp}, where the callee’s code is executed even before the bundle is accepted; 
    \item \textsf{executeUserOp}, which performs arbitrary low-level calls as long as gas sponsorship conditions are met; 
    \item pre-verification gas sponsorship via \textsf{Paymasters}, whose logic may trigger secondary calls into attacker-selected targets.
\end{enumerate}
In all cases, the call is indistinguishable from routine ERC-4337 execution, creating a trigger mechanism for the attacker that is automated, silent, and entirely decoupled from the victim’s actions.

\smallskip
\noindent\textbf{Stateful multi-stage exploitation.}
The combination further enables multi-round exploitation patterns. A single EIP-7702 authorization tuple installs a durable delegation. ERC-4337 then provides unbounded opportunities for activation: any future \textsf{UserOperation}, even those sent by third parties, may trigger the delegate logic. As a result, the attacker can:
\begin{itemize}
    \item embed the delegation at time $t_0$ (phishing phase),
    \item wait indefinitely,
    \item trigger execution at arbitrary later times $\{t_1, t_2, \dots\}$ using crafted \textsf{UserOperations},
    \item or rely on ambient 4337 wallet activity by unrelated users to trigger the delegate.
\end{itemize}

\noindent\textbf{Consequences for the 4337 Ecosystem.}
Once a 7702 delegation is installed, the resulting interactions with ERC-4337 extend the impact of a single-account compromise into the broader account-abstraction pipeline. (i) Because bundlers unconditionally relay \textsf{UserOperations} that target an address, they can be unintentionally transformed into activation carriers, repeatedly invoking the delegated victim account. (ii) \textsf{Paymasters} further remove the economic barrier to such activations by covering gas fees, enabling an attacker to issue unlimited trigger attempts at negligible cost. (iii) At the protocol layer, the \textsf{EntryPoint} executes both validation and execution steps against the delegated victim address, thereby running attacker-controlled logic under the guise of a normal user-operation flow. Meanwhile, most 4337 wallet interfaces do not surface changes to an EOA’s code field, rendering the presence of a delegation invisible to the user. 

% These factors collectively convert a localized phishing compromise into a system-level vulnerability pathway, where core 4337 components unknowingly amplify and sustain delegated execution.

\subsection{Case Study: In-the-Wild Composite Attack}
\begin{figure}[t]
    \centering
    \includegraphics[width=\linewidth]{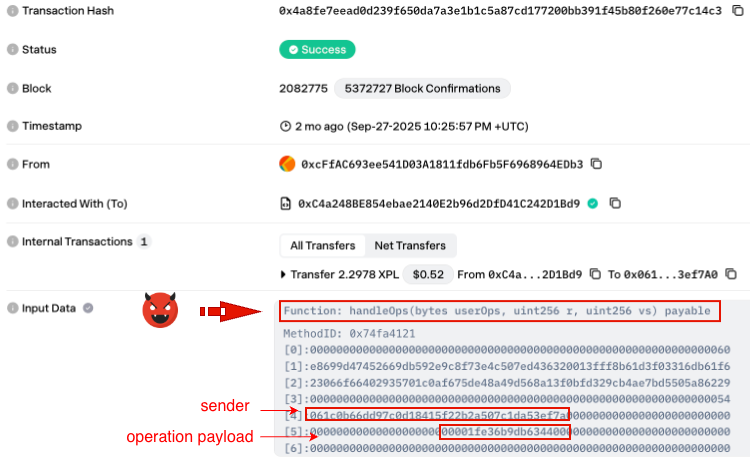}
    \vspace{-0.1in}
    \caption{\textbf{In-the-wild EIP-7702 × ERC-4337 composite invocation.}
    The transaction (\texttt{0x4a8fe7ee...c14c3}) is a standard ERC-4337 
    \textsf{handleOps} call, not a typed 7702 \textsf{AuthTx}.  
    Nevertheless, execution is routed into delegated code because the 
    callee (\texttt{0xC4a248BE...D1Bd9}) carries a 7702-style delegation indicator.}
    \vspace{-0.05in}
    \label{fig:4337-composite}
\end{figure}

We further analyze a real-world transaction observed on the PlasmaScan blockchain explorer (Tx: \textit{\href{https://plasmascan.to/tx/0x4a8fe7eead0d239f650da7a3e1b1c5a87cd177200bb391f45b80f260e77c14c3}{0x4a8fee...c14c3}}) (Fig.~\ref{fig:4337-composite}). 
Although the transaction is not an EIP-7702 \textsf{AuthTx} (type~0x04), its execution exhibits a precise interaction pattern between delegated EOA execution (7702-style) and ERC-4337’s \textsf{handleOps} flow, illustrating how bundlers can unintentionally deliver attacker-crafted payloads to an already delegated EOA.

\smallskip
\noindent\textbf{Execution context.}
The transaction originates from \textit{0xcFf…64EDb3} and targets \textit{0xC4a…D1Bd9}. 
Blockchain traces reveal that this target is not a conventional contract wallet but an EOA whose code segment contains a 7702-style delegation indicator (\texttt{0xef0100 || delegate}), turning the account into a "pseudo-contract" whose control flow is redirected to attacker-supplied logic.
Despite this, the transaction invokes the ERC-4337 interface:
\begin{quote}
\textsf{handleOps(bytes userOps, uint256 r, uint256 vs)}  
\end{quote}
with MethodID \texttt{0x74fa4121}, consistent with \textsf{EntryPoint}~0.6.0 deployments.

% \smallskip
% \noindent\textbf{Structure of the composite invocation.}
The calldata contains seven ABI-decoded fields, of which the following two are most significant:
\begin{enumerate}[label=(\alph*)]
    \item The element at slot~\texttt{[4]},  
    \texttt{0x061c0b66dd97c0d1841...},  
    corresponding to the \textsf{sender} or called account within the \textsf{userOp}.  
    This address is subsequently the recipient of internal calls emitted by the EntryPoint.
    \item The element at slot~\texttt{[5]} (\texttt{0x00...1fe36b9db6344...}) encodes an operation payload routed to the delegated EOA. Although it is not an AuthTx, the payload meets the minimal requirements for execution: \textit{any} externally triggered call to the victim account is redirected to the installed delegate.
\end{enumerate}

\noindent\textbf{Observed behavior.}
The execution trace confirms one internal transaction:
\texttt{From: 0xC4a2...D1Bd9}  
\quad\texttt{To: 0x061c...3ef7A0}
with a transferred value of \textit{1 wei}.  
Such an internal call cannot occur for a native EOA; it is only possible because the EOA has been rewritten through a 7702-style delegation indicator. Once the \textsf{EntryPoint} delivers the \textsf{handleOps} payload, the EOA’s control flow is immediately redirected into the delegate contract, consistent with the delegated-execution semantics.

\noindent\textbf{Implications.}
This transaction provides empirical evidence that:
\begin{packeditemize}
    \item ERC-4337’s \textsf{handleOps} can serve as an unsolicited activation vector for accounts modified by EIP-7702 delegation.
    \item Bundlers and relayers, which execute \textsf{handleOps} as part of ordinary user-operation flows, may unknowingly relay attacker-crafted payloads into delegated EOAs.  
    \item The delegated EOA is treated by the \textsf{EntryPoint} like a contract wallet, enabling arbitrary execution of the delegate’s logic without requiring a 7702 transaction or additional victim signatures.
\end{packeditemize}

The case shows that once a victim's account is rewritten, the ERC-4337 infrastructure itself, in particular \textsf{bundler}s and \textsf{EntryPoint}s, becomes an involuntary relayer of attacker-selected execution.

\subsection{\textcolor{magenta}{(b)} Replay-like Cross-chain Exploits}
EIP-7702 introduces a special authorization mode in which the tuple’s \texttt{chainId} is set to \textbf{zero}, signaling that the signature is chain-agnostic and should validate on any network. While intended for cross-rollup or L2 migration workflows, this design unintentionally reintroduces a \textbf{replay-like} threat surface: a single phishing signature can propagate across multiple chains, each interpreting the authorization tuple as locally valid and independently installing the delegation indicator.

The resulting failure mode is qualitatively different from classical transaction replay. Traditional replay exploits reuse the \emph{same} transaction in multiple domains; in contrast, a 7702 tuple with \texttt{chainId = 0} enables the attacker to craft \emph{fresh} AuthTx transactions on arbitrary networks, all anchored in the victim’s original signature. The attacker can therefore reproduce the compromise across L1, rollups, testnets, or application-specific L2s, even if the user interacts with only one domain. Worse, nonce constraints do not provide meaningful containment: each chain maintains its own independent EOA nonce, so the same tuple can pass validation simultaneously on every network where the victim address exists.

This transforms a single-signature phishing event into a multi-domain takeover. An attacker who installs a malicious delegate on one chain can subsequently replicate the installation on any other chain, drain assets native to each domain, and persist across bridges and sequencer boundaries. In multi-chain wallet ecosystems, where users routinely reuse the same EOA across L1, L2s, and sidechains, the \texttt{chainId = 0} mode creates a systemic amplification vector: compromise in one environment silently metastasizes into a cross-chain compromise without requiring additional user actions. This cross-domain persistence fundamentally alters the blast radius of 7702-based phishing, turning what appears to be a localized attack into a platform-wide exploitation channel.

\begin{table}[!ht]
\centering
\small
\renewcommand{\arraystretch}{1}
\setlength{\tabcolsep}{1pt}
\caption{\textbf{Cross-chain Replay Attack Outcomes.} 
All three chains independently accept the same \texttt{chainId=0} authorization tuple.}
\label{tab:3chain-single}
\begin{tabular}{c|cc|cc}
\toprule
\textbf{Chain (Chain ID)} &
\multicolumn{2}{c|}{\textbf{ETH (V $\rightarrow$ A)}} &
\multicolumn{2}{c}{\textbf{ERC-20 (V $\rightarrow$ A)}} \\
\cmidrule{2-5}
 & \textbf{Before} & \textbf{After} & \textbf{Before} & \textbf{After} \\
\midrule

Chain A (1337) & 10000.0 & 0.000432931 & 2000.0 & 0.0 \\
Chain B (2337) & 10000.0 & 0.000432931 & 2000.0 & 0.0 \\
Chain C (3337) & 10000.0 & 0.000432931 & 2000.0 & 0.0 \\

\midrule
\textbf{Total} & \textbf{30000.0\,\ethsym} & \textemdash & \textbf{6000.0\,\ercsym} & \textemdash \\
\bottomrule
\end{tabular}
\end{table}

\vspace{-0.1in}
\subsection{Cross-chain Exploits Experiment}
To empirically validate whether a single EIP-7702 authorization tuple with \texttt{chainId = 0} can be replayed across multiple execution environments, we constructed a controlled three-chain environment using independent Hardhat instances (chain IDs 1337, 2337, and 3337), all sharing the same EOA keyspace but maintaining fully disjoint nonce and state domains. A malicious delegate contract was deployed on each chain at an identical address, mirroring the conditions under which cross-chain wallet reuse occurs in practice. The victim then signed a chain-agnostic authorization tuple (\texttt{chainId = 0}), after which the attacker submitted the \emph{same} tuple unmodified to all three chains.

Across all domains, the tuple validated successfully: each chain independently verified the victim’s signature, found the victim’s nonce to match the advertised tuple nonce, and installed the malicious delegation code \texttt{0xef0100\,$\|$\,target}. As a consequence, a routine self-call from the victim on each chain immediately executed the fallback logic of the attacker’s delegate, draining all ETH and ERC-20 balances into the attacker-controlled sink address. The replay succeeded identically despite the chains having distinct block histories and gas markets, demonstrating that no inter-chain coordination or state sharing is required for exploitation.

Experiment results confirm that the \emph{chain-agnostic authorization mode} creates a true multi-domain replay surface: a single phishing signature can be used to compromise every chain on which the victim reuses the same EOA, with each chain treating the authorization as fresh and locally valid. For more details, see Appendix~\ref{appendix:crosschain-details}.

\section{Threat Mitigation}
\label{sec-mitigation}
%================================================= 

\subsection{Protocol-level Constraints on Delegation}
The phishing vectors we identify stem from structural design choices in EIP-7702 rather than from incidental implementation issues. The delegation mechanism grants EOAs the ability to overwrite their account code with an arbitrary delegate target, yet the protocol places almost no constraints on what that target may do or how the delegation may subsequently be triggered. This openness fundamentally expands the attack surface.

\smallskip
\noindent\textit{\ding{172} Unscoped delegation semantics.}
At the semantic level, an authorization tuple carries only \texttt{(chainId, target, nonce, yParity, r, s)} and encodes no execution scope, call restrictions, expiry, or safety invariants. Once accepted, the EOA permanently routes \emph{all} calls, e.g., self-calls, third-party calls, protocol callbacks, and even ERC-4337 \textsf{entrypoint} calls, into the delegate’s fallback logic. As our experiments show, this global binding makes a single phishing signature sufficient to induce total execution takeover.

% \smallskip
\noindent\textit{\ding{173} Chain-agnostic authorization replay.}
A second source of risk arises from the tuple’s optional chain-agnostic mode (\texttt{chainId = 0}). Because EOA nonces evolve independently on each chain, the same signature can be replayed across multiple domains, allowing attackers to replicate delegation installation on every network where the victim address exists. We demonstrate that this mechanism directly enables cross-chain metastasis of a single phishing event.

\smallskip
\noindent\textbf{Mitigation strategies.}
\ding{172} A mitigation is to require authorization tuples to carry explicit execution scope. Clients should reject unscoped delegations and instead mandate constraints such as expiry bounds or single-use semantics. Wallets can additionally enforce "foreground-only" delegation\footnote{"Foreground-only" delegation restricts delegated execution to transactions explicitly initiated by the user through the wallet UI.}, preventing automatic activation by protocol callbacks or external contracts. \ding{173} To neutralize cross-chain replay, implementations should forbid \texttt{chainId = 0} or treat it as invalid unless accompanied by context-binding metadata (e.g., chain-specific salts). Wallets can implement preflight checks that compare the tuple’s declared chain with the current network, while nodes reject tuples whose chain field does not match the \texttt{chainId}.

\subsection{4337 Ecosystem Defenses}
\noindent\textbf{Threats within the ERC-4337 pipeline.}
EIP-7702 delegation interacts with the ERC-4337 execution pipeline in a structurally hazardous way. Once a malicious delegate is installed, any subsequent \textsf{UserOperation} that targets the victim’s address, even indirectly through a router contract, a paymaster pre-verification callback, or another wallet’s batched operation, causes the \textsf{EntryPoint} to route control into attacker-supplied code. Because bundlers generally treat \textsf{validateUserOp} as pure verification rather than execution, the delegated fallback can be activated before the bundle is accepted, shifting phishing activation from a user-driven event to a network-driven one. Paymasters further expand the attack surface: if a malicious or compromised paymaster sponsors gas for a crafted \textsf{UserOp}, the attacker can trigger delegated execution without paying gas and without victim interaction. Finally, since ERC-4337 wallets often abstract away raw calldata and signature details, UIs are poorly positioned to surface delegation state, leaving users unaware that  \textsf{EntryPoint} will repeatedly and automatically relaunch attacker-controlled execution on their behalf.

\smallskip
\noindent\textbf{Mitigation strategies.}
A principled defense requires filtering delegated execution at the \textsf{EntryPoint}, the bundler mempool, and the paymaster validation path. \ding{172} Bundlers should perform pre-simulation under the assumption of potential delegation and reject \textsf{UserOps} whose top-level target has non-empty EOA code matching the 7702 delegation prefix. \ding{173} The \textsf{EntryPoint} can restrict delegated execution in \textsf{validateUserOp} by enforcing a static check: EOAs with installed delegation must not be treated as wallets unless they explicitly implement the expected interface. \ding{174} Paymasters should refuse to fund \textsf{UserOps} that indirectly route control to an EOA with active delegation, eliminating gas-subsidized activation vectors. \ding{175} Wallet UIs must surface delegation status and warn when 4337 pipelines would dispatch through attacker-controlled code. 

%=================================================
\section{Related Work}
\label{sec-rw}
%================================================

\noindent\textbf{Ethereum security issues.} 
Ethereum’s attack surface spans contract correctness, inter-contract interaction, and system-level protocol behaviors. Early efforts characterize the fragility of upgradeable smart contracts and proxy patterns, revealing widespread misconfigurations that expose privilege-escalation and contract-hijacking risks~\cite{li2024characterizing}, while complementary program-analysis systems strengthen on-chain correctness through bytecode rewriting for instant patching~\cite{rodler2021evmpatch} or formal inference of financial-security invariants~\cite{wang2023automated}. Cross-contract communication further introduces confused-deputy vulnerabilities that arise even when individual modules are correct~\cite{gritti2023confusum}, and large-scale forensic systems reconstruct historical EVM traces to detect reentrancy, unchecked calls, and suicidal contracts directly from transactions~\cite{zhang2020txspector}. 

Beyond functional correctness, newer measurement studies uncover systemic weaknesses in address-poisoning scams exploiting UI truncation in wallets~\cite{guan2024characterizing}, in token ecosystems where disposable assets and sniper bots enable large-scale rug-pull extraction~\cite{cernera2023token}, and in coordinated phishing gangs whose transaction motifs reveal complex criminal networks~\cite{liu2024fishing}.

% System-layer vulnerabilities further widen Ethereum’s security boundary. Mempool-focused work reveals asymmetric DoS vectors, including stealthy eviction, locking, and speculative resource exhaustion, exploiting the need to execute Turing-complete transactions before inclusion~\cite{wang2024understanding,yaish2024speculative}. Network- and consensus-adjacent analyses show that Ethereum clients can be coerced during initial sync to adopt adversarially crafted forks~\cite{taverna2023snapping}, while validators can be deanonymized by correlating peer-level metadata in the P2P network~\cite{heimbach2025deanonymizing}. On top of the base layer, rollup ecosystems inherit and amplify liveness risks, as adversaries craft malformed yet legality-bypassing transactions that exhaust sequencer resources at zero cost~\cite{li2025denial}. Meanwhile, the phishing landscape has industrialised: large-scale studies reveal Drainer-as-a-Service economies~\cite{he2025unmasking} and tens of thousands of autonomous phishing contracts executing hundreds of thousands of fraudulent transactions~\cite{he2025phishing}.

\smallskip
\noindent\textbf{Blockchain phishing attacks (Table~\ref{tab:phishing_attack_comparison}).} 
Blockchain phishing has expanded from classical user deception to multi-layer manipulation of signatures, calldata, and contract semantics. Early attacks rely on spoofed DApps, misleading wallet-connect interfaces, and impersonated administrators to trick users into signing harmful transactions~\cite{he2023txphishscope,idehen2024web3,chainalysis2023midyear}, while airdrop/giveaway campaigns weaponize social-engineering incentives for credential harvesting~\cite{li2023double,filiba2024scams,slowmist2025analysis}. As phishing techniques matured, adversaries shifted toward transaction-level manipulation: crafted calldata and deceptive previews conceal token-drain behavior~\cite{chen2024dissecting,pshanichnaya2024solana,cholakov2025drainers}, and approval-forging attacks exploit users’ misunderstanding of ERC-20 allowance semantics to gain unbounded control over assets~\cite{microsoft2022icephish,chainalysis2023approval}. 

% These works collectively show that phishing no longer depends solely on social trust exploitation but increasingly leverages mismatches between user-visible interfaces and on-chain execution.

More recent attacks embed phishing logic directly into smart contracts and application. Malicious NFT minting and marketplace approval traps~\cite{wang2023account} abuse callbacks or \texttt{setApprovalForAll} to seize assets; upgradeable proxies enable covert logic substitution; and typed-data (\texttt{permit}) signatures allow replayable or effectively unlimited approvals under benign-looking prompts. Ecosystem-specific variants further target DeFi routers via impersonated swap or liquidity flows, or manipulate MEV bots with false arbitrage signals that trigger draining callbacks. Phishing has become a cross-layer paradigm that exploits calldata semantics and typed signatures, setting the stage for our identification of EIP-7702 authorization hijacking as a new form of delegated-execution phishing.

\smallskip
\noindent\textbf{Ethereum phishing detection}.
A substantial body of research models phishing detection on Ethereum as a graph- or sequence-classification task over the transaction network. Early approaches apply network embeddings or GNN-based feature extraction to capture structural and statistical characteristics of accounts~\cite{wu2020phishers,chen2020phishing}. Subsequent works enhance representation learning through temporal modeling, heterogeneous transaction sequences, and subgraph-centric abstractions that better encode higher-order dependencies and alleviate sparsity~\cite{lin2023tracking,li2022ttagn,wang2025tsgn}. More recent models further integrate adaptive attention, dynamic graph evolution, and self-supervised or incremental learning to improve robustness to long-tail degree distributions and rapidly changing on-chain activity~\cite{sun2024adaptive,zhang2024grabphisher,li2023siege}. These methods collectively show that structural signals, temporal trajectories, and localized transaction motifs are informative for identifying phishing accounts in large-scale networks.

A complementary line of work focuses on behavioral signals and phishing vectors. Temporal-dependence models capture both stable and fluctuating user behavior patterns to distinguish deceptive accounts in dense and noisy environments~\cite{ghosh2025temper,ghosh2025catalog}. Contrastive learning and architecture-search frameworks further improve generalization in label-scarce settings~\cite{li2023tgc,liu2025phishing}. Beyond address-centric detection, recent empirical studies~\cite{he2023txphishscope,chen2024dissecting,yang2024stole} characterize specialized scams, such as transaction-based phishing, payload-manipulation attacks, and NFT-focused theft, revealing previously overlooked attack surfaces and releasing large-scale ground-truth datasets. These works demonstrate Ethereum phishing from simple financial-transfer scams to complex contract-level deception, but they largely assume trustworthy transaction semantics and thus do not account for \emph{authorization-layer} abuses central to our threat model.

%=================================================
\section{Conclusion}
\label{sec-conclusion}
%================================================
We reveal that EIP-7702 creates an attack surface fundamentally different from traditional transaction phishing: a shift from \emph{transaction-level consent} to \emph{account-level takeover}. We formally characterize the mechanism, demonstrate its exploitability, and measure its real-world footprint. We demonstrate that a single phishing signature can induce durable and remotely activatable compromise both practically (attack under control) and empirically (via large-scale on-chain measurement). 
We further uncover compositional attack surfaces (with ERC-4337 and multi-chain) and outline practical mitigations for both protocol designers and wallet implementers.

%================================================
\bibliographystyle{unsrt}
\bibliography{bib.bib}
%================================================

\begin{table*}[t]
\renewcommand{\arraystretch}{1}
\centering
\caption{Abbreviations (i.e., \textcolor{teal}{type.id}) for entities used in our EIP-7702 phishing and cross-chain replay experiments.}
\label{tab:abbr-7702}
\resizebox{0.99\textwidth}{!}{
\begin{tabular}{cc|cc|c}
\toprule
\multicolumn{2}{c}{\textbf{Type}} &
\textbf{ID} &
\textbf{Full Identifier} &
\textbf{Notes} \\
\midrule

%%%%%%%%%%%%%%%%%%%%%%%%%%%%%%%%%%%%%%%%%%%%%%%%%%%%%%
% ADDRESSES
%%%%%%%%%%%%%%%%%%%%%%%%%%%%%%%%%%%%%%%%%%%%%%%%%%%%%%
\multirow{6}{*}{\rotatebox{90}{Address}}   
& \textcolor{teal}{adr.} & \textcolor{teal}{2266} 
& 0xf39Fd6e51aad88F6F4ce6aB8827279cffFb92266 
& Victim EOA (same across all 3 chains) \\

& \textcolor{teal}{adr.} & \textcolor{teal}{79c8} 
& 0x70997970C51812dc3A010C7d01b50e0d17dc79C8 
& Attacker EOA / sink (same across chains) \\

& \textcolor{teal}{adr.} & \textcolor{teal}{0aa3} 
& 0x5fbdb2315678afecb367f032d93f642f64180aa3 
& Malicious delegate (local experiment) \\

& \textcolor{teal}{adr.} & \textcolor{teal}{18bc} 
& 0x8464135c8F25Da09e49BC8782676a84730C318bC 
& Malicious delegate (cross-chain attack) \\

& \textcolor{teal}{adr.} & \textcolor{teal}{3292} 
& 0x71C95911E9a5D330f4D621842EC243EE1343292e 
& Mock ERC-20 token (same address on all chains) \\

& \textcolor{teal}{adr.} & \textcolor{teal}{6602} 
& 0x663f3ad617193148711d28f5334ee4ed07016602 
& Dummy protocol contract (ambient trigger) \\

\cmidrule{1-3}

%%%%%%%%%%%%%%%%%%%%%%%%%%%%%%%%%%%%%%%%%%%%%%%%%%%%%%
% TRANSACTIONS — LOCAL EXPERIMENT
%%%%%%%%%%%%%%%%%%%%%%%%%%%%%%%%%%%%%%%%%%%%%%%%%%%%%%
\multirow{10}{*}{\rotatebox{90}{Transaction}}  
& \textcolor{teal}{txid} & \textcolor{teal}{7d62} 
& 0x6b78de5fb9d70f8a9f1ade14cd52a11f6093493c76489f31c4d14b6341307d62 
& AuthTx installing delegation (local) \\

& \textcolor{teal}{txid} & \textcolor{teal}{13a0} 
& 0x2d2346629372f11a684d712d3024b71576d2666ef655c49003d6940b68d613a0 
& Deployment of malicious delegate \\

& \textcolor{teal}{txid} & \textcolor{teal}{b64f} 
& 0x2b6d3367fea34d9f3b2475e622279139d422cda7ff7c18530837508e4b1eb64f 
& Scenario A — user-driven trigger \\

& \textcolor{teal}{txid} & \textcolor{teal}{bb1f} 
& 0x5d570a6a057fe11091efca68757c81a59f67cacb140db4a7db1cce8e5730bb1f 
& Scenario B — attacker-driven trigger \\

& \textcolor{teal}{txid} & \textcolor{teal}{b0c1} 
& 0xb616a70460da2a0f7a113bde6fafede063d35e7d7c4255311f7b4ffab9a3b0c1 
& Scenario C — ambient trigger \\

\cmidrule{2-3}

%%%%%%%%%%%%%%%%%%%%%%%%%%%%%%%%%%%%%%%%%%%%%%%%%%%%%%
% TRANSACTIONS — CROSS-CHAIN (AUTHORIZATION)
%%%%%%%%%%%%%%%%%%%%%%%%%%%%%%%%%%%%%%%%%%%%%%%%%%%%%%
& \textcolor{teal}{txid} & \textcolor{teal}{A.auth} 
& 0x53dd7567680e408687d2a8fc9d9f2893474c0d068bdaf666b2e95999cdc976bc 
& Chain A authorization (\texttt{chainId=0}) \\

& \textcolor{teal}{txid} & \textcolor{teal}{B.auth} 
& 0xbc3b443e09f3130a63250cd5191515fdc0fa30faf3957cf12ecdff08e59e0595 
& Chain B authorization \\

& \textcolor{teal}{txid} & \textcolor{teal}{C.auth} 
& 0x6461ee84b806e053cd22809dedb11fc9ae621c86362ea59f8d36d73c19ab25a1 
& Chain C authorization \\

%%%%%%%%%%%%%%%%%%%%%%%%%%%%%%%%%%%%%%%%%%%%%%%%%%%%%%
% TRANSACTIONS — CROSS-CHAIN (TRIGGER)
%%%%%%%%%%%%%%%%%%%%%%%%%%%%%%%%%%%%%%%%%%%%%%%%%%%%%%

& \textcolor{teal}{txid} & \textcolor{teal}{A.trg} 
& 0xbe2f31a4519c97332c3c3f97ccd248c8b3bc1ced43ef9f1a955c46c329b0fac4 
& Chain A trigger transaction \\

& \textcolor{teal}{txid} & \textcolor{teal}{B.trg} 
& 0x3b955c7e7f6fc6e2ea66fdf93233a1f1b081723b6701eb51fca0f72cb57954dc 
& Chain B trigger transaction \\

& \textcolor{teal}{txid} & \textcolor{teal}{C.trg} 
& 0x89ae46d7c694707865b1ebcf7c4a2314782fa9bdd72276b92535c1352fa3c35b 
& Chain C trigger transaction \\

\cmidrule{1-3}

% %%%%%%%%%%%%%%%%%%%%%%%%%%%%%%%%%%%%%%%%%%%%%%%%%%%%%%
% % BLOCKS
% %%%%%%%%%%%%%%%%%%%%%%%%%%%%%%%%%%%%%%%%%%%%%%%%%%%%%%
% \multirow{3}{*}{\rotatebox{90}{Block}}
% & \textcolor{teal}{blk.} & \textcolor{teal}{0001} 
% & 0x1 
% & Block of delegate deployment \\

% & \textcolor{teal}{blk.} & \textcolor{teal}{A.blk} 
% & 0x1 
% & Chain A authorization block \\

% & \textcolor{teal}{blk.} & \textcolor{teal}{B.blk} 
% & 0x1 
% & Chain B authorization block \\

% & \textcolor{teal}{blk.} & \textcolor{teal}{C.blk} 
% & 0x1 
% & Chain C authorization block \\

% \cmidrule{1-3}

%%%%%%%%%%%%%%%%%%%%%%%%%%%%%%%%%%%%%%%%%%%%%%%%%%%%%%
% 7702 TUPLES & DELEGATION
%%%%%%%%%%%%%%%%%%%%%%%%%%%%%%%%%%%%%%%%%%%%%%%%%%%%%%
\multirow{4}{*}{\rotatebox{90}{7702}}  
& \textcolor{teal}{auth.} & \textcolor{teal}{msg} 
& MAGIC || RLP([chainId=0,\ target=0x8464...18bc,\ nonce=0]) 
& Chain-agnostic signing message \\

& \textcolor{teal}{auth.} & \textcolor{teal}{sig.}
& r = 0x9a9a1b..., s = 0x67668..., yParity = 1
& Signature extracted from tuple \\

& \textcolor{teal}{del.} & \textcolor{teal}{ef01} 
& 0xef01008464135c8f25da09e49bc8782676a84730c318bc 
& Delegation code written on all 3 chains \\

& \textcolor{teal}{tuple} & \textcolor{teal}{chain0} 
& 0xf85a0094846413...b2c6166 
& RLP-encoded \texttt{chainId=0} tuple \\

\bottomrule
\end{tabular}
}
\end{table*}

\appendix

\section*{Ethical Considerations}

This work studies a protocol-level phishing vector enabled by EIP-7702. We aim to understand and mitigate systemic security risks, not to facilitate exploitation. All attacks are analyzed from a defensive perspective, with the intent of revealing previously unexamined failure modes in delegated execution and informing safer protocol and wallet designs. We do not advocate the deployment of any techniques in real-world environments.

Our experiments were conducted exclusively in controlled settings, including local test environments and post-hoc analysis of publicly available on-chain data. We do not interact with real users, solicit signatures, or induce delegations on live accounts. Empirical measurements rely solely on historical blockchain data and publicly observable transactions, without deanonymization attempts of off-chain identities. No private keys, credentials, or sensitive personal information were accessed or collected at any stage.

To reduce the risk of misuse, we  omit exploit-ready artifacts, such as deployable malicious contract templates or end-to-end phishing scripts. Where attack mechanisms are discussed, they are described at a conceptual and analytical level sufficient for security understanding but not optimized for replication. We have shared our findings with relevant ecosystem stakeholders and focus our contributions on protocol-level and ecosystem-level mitigations.

\section{Full Experimental Details}
\label{appendix:crosschain-details}

\subsection{Multi-chain Environment Setup}
We instantiated three independent Hardhat networks on local ports \texttt{8545}, \texttt{9545}, and \texttt{10545}, configured with chain identifiers \texttt{1337}, \texttt{2337}, and \texttt{3337}, respectively. All networks shared the same BIP-39 mnemonic:
\vspace{-0.12in}
\begin{lstlisting}[language=text]
test test test test test test test test test test test junk
\end{lstlisting}
\vspace{-0.12in}
This ensured that the victim (\textcolor{teal}{adr.2266}) and attacker (\textcolor{teal}{adr.79c8}) EOAs appeared identically across all three chains, while maintaining independent nonce spaces and account states.

\smallskip
\noindent\textbf{Node configuration.}
\vspace{-0.12in}
\begin{lstlisting}[language=javascript]
networks: {
  chainA: { url: "http://127.0.0.1:8545",  chainId: 1337, accounts: { mnemonic } },
  chainB: { url: "http://127.0.0.1:9545",  chainId: 2337, accounts: { mnemonic } },
  chainC: { url: "http://127.0.0.1:10545", chainId: 3337, accounts: { mnemonic } },
}
\end{lstlisting}
\vspace{-0.12in}

\subsection{Uniform Deployment of the Delegate Contract}

A malicious delegate contract (\textcolor{teal}{adr.18bc}) was deployed on all three chains using the attacker’s EOA. Because the same mnemonic and identical deployment nonce were used, all deployments resolved deterministically to the same address:
\vspace{-0.12in}
\begin{lstlisting}[language=text]
   0x8464135c8F25Da09e49BC8782676a84730C318bC
\end{lstlisting}
\vspace{-0.12in}

\subsection{Construction of Chain-agnostic Authorization Tuple}

We created a single (7702) authorization tuple with \texttt{chainId = 0}, instructing the victim to delegate execution to a malicious contract:

\begin{lstlisting}[language=text]
  0xf85a00948464135c8f25da09e49bc8782676a84730c31
  8bc0001a09a9a1bd58376d5185d421b67c5c76078cd7d74
  70b27987faa519fb3015f7df3ca0676682acce38380ea0c
  e9c4a2683841c01906ea83e156466e47310805b2c6166
\end{lstlisting}
\vspace{-0.12in}
Decoded tuple components:
\begin{itemize}
  \item \textbf{chainId}: 0 (chain-agnostic)
  \item \textbf{target}: \textcolor{teal}{adr.18bc}
  \item \textbf{nonce}: 0
  \item \textbf{signature}: $(y,r,s)$ matching victim’s key
\end{itemize}

This tuple was saved as \texttt{tuple\_chain0.hex} and reused verbatim across all three networks.

\begin{figure*}[t]
\centering

% ---- 左图 ----
\subfigure[Top 10 ETH Victims by Loss]{%
    \includegraphics[width=0.32\textwidth]{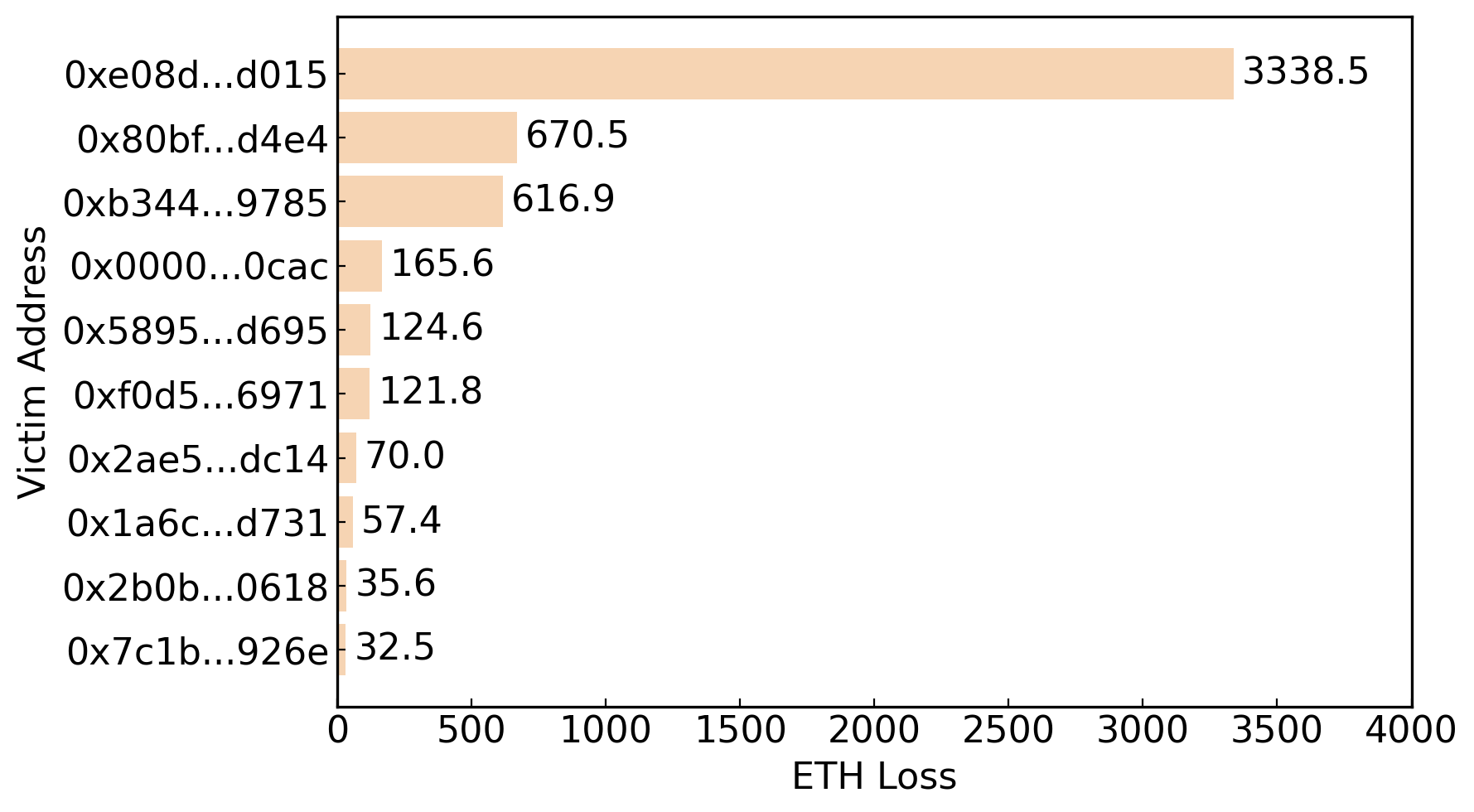}
    \centering
    \label{fig:top-eth-victims}
}
\hfill
% ---- 中图 ----
\subfigure[Top 10 ETH Delegators by Total Loss]{%
    \includegraphics[width=0.32\textwidth]{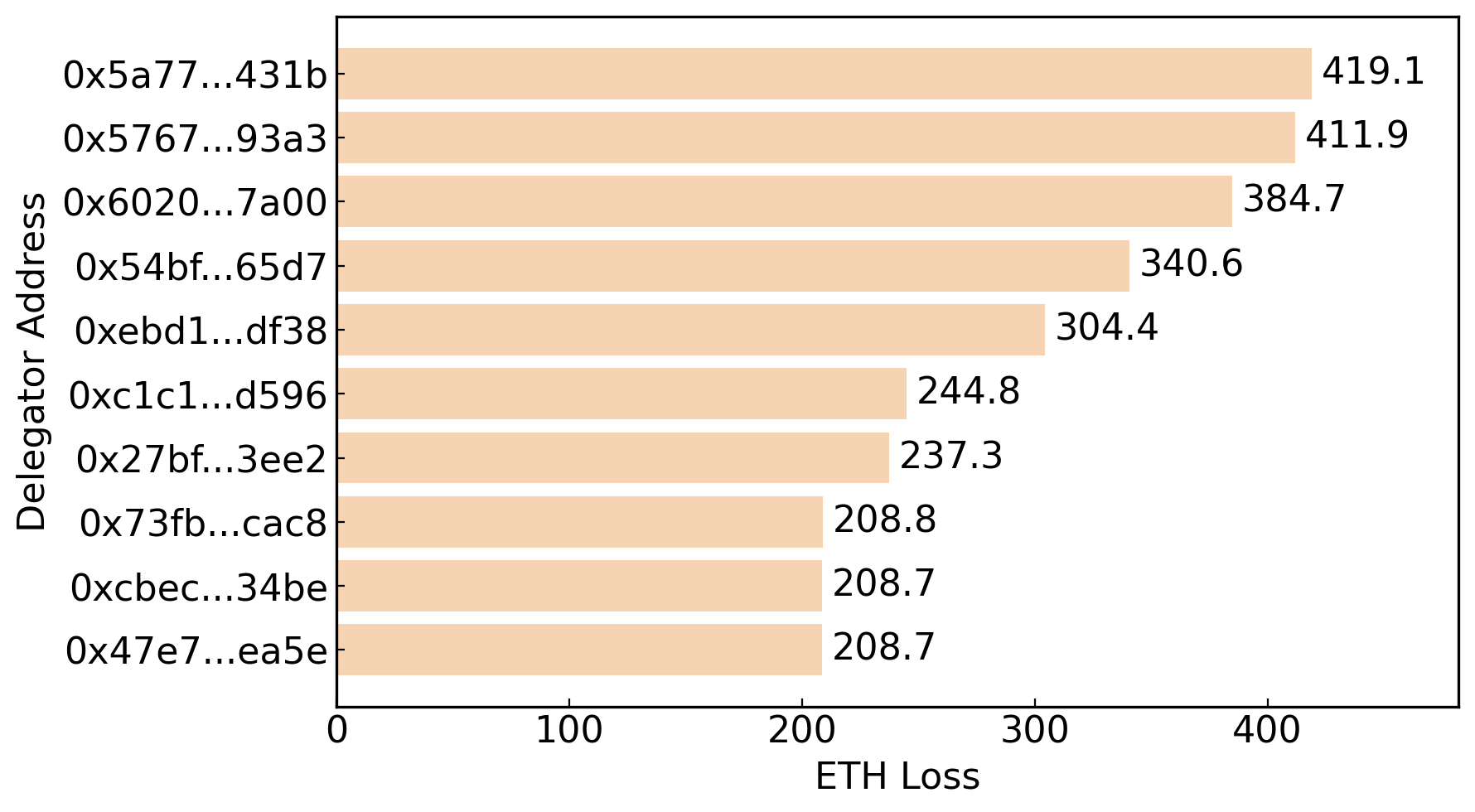}
    \centering
    \label{fig:top-eth-delegators}
}
\hfill
% ---- 右图 ----
\subfigure[Top 10 NFT Contracts by Stolen Items]{%
    \includegraphics[width=0.32\textwidth]{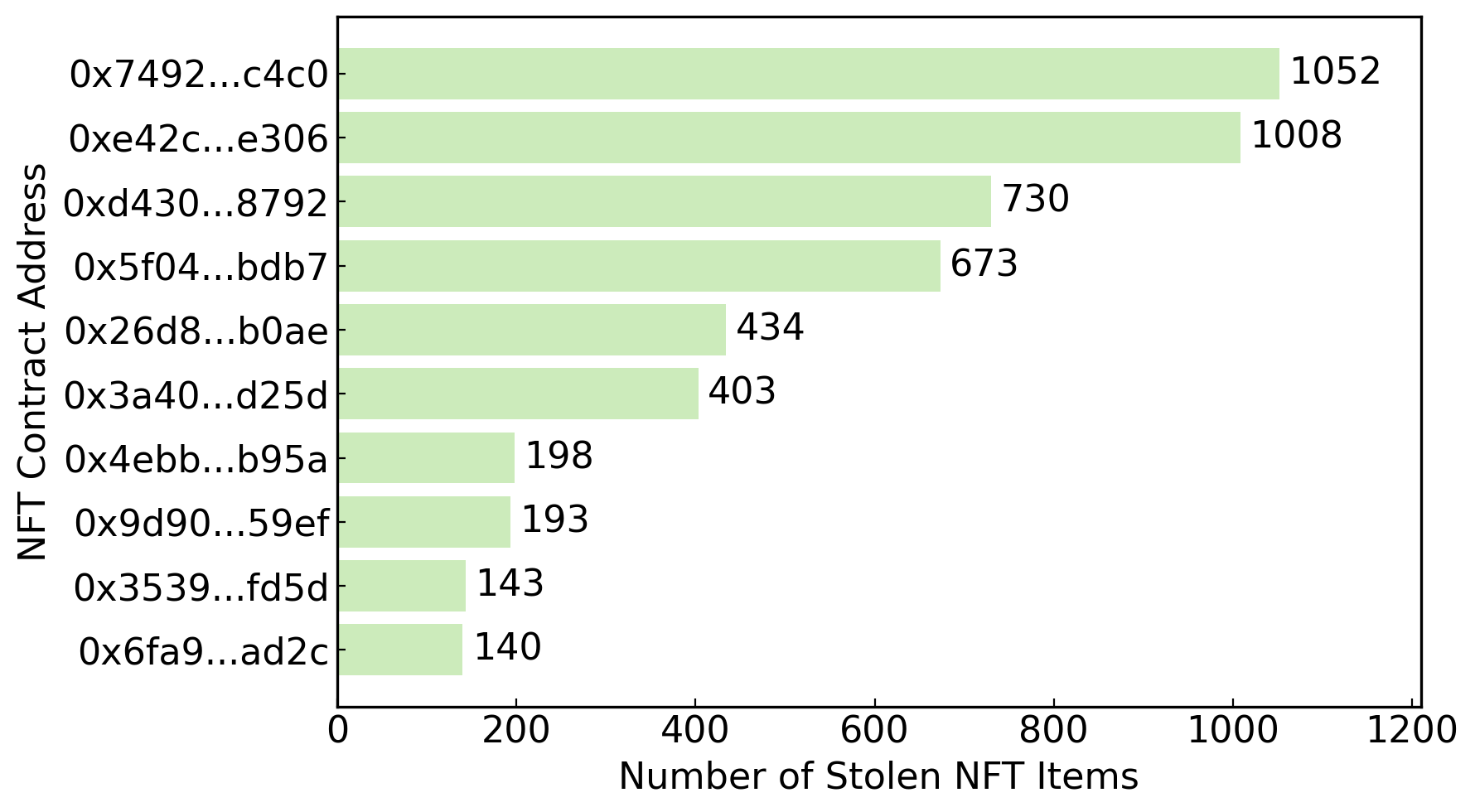}
    \centering
    \label{fig:top-nft-stolen}
}

\caption{Cross-category comparison of high-loss entities in EIP-7702–related incidents.}
% \label{fig:loss-landscape}
\end{figure*}

\subsection{Replaying the Authorization on Three Chains}

The authorization tuple was submitted to all networks through identical \texttt{AuthTx} transactions.

% \smallskip
\noindent\textbf{Chain A authorization.}
\vspace{-0.12in}
\begin{lstlisting}[language=text]
txHash = 0x53dd7567680e4086...e95999cdc976bc
victim.code = ef0100 8464135c8...a4730c318bc
\end{lstlisting}
\vspace{-0.12in}

% \smallskip
\noindent\textbf{Chain B authorization.}
\vspace{-0.12in}
\begin{lstlisting}[language=text]
txHash = 0xbc3b443e0...57cf12ecdff08e59e0595
victim.code = ef0100 8464135c8...a4730c318bc
\end{lstlisting}
\vspace{-0.12in}

% \smallskip
\noindent\textbf{Chain C authorization.}
\vspace{-0.12in}
\begin{lstlisting}[language=text]
txHash = 0x6461ee84b806e0...8d36d73c19ab25a1
victim.code = ef0100 8464135c8...84730c318bc
\end{lstlisting}
\vspace{-0.12in}
In all cases, the tuple was accepted despite originating from a different chain context, confirming that \texttt{chainId = 0} bypasses chain-specific authorization checks.

\subsection{Triggering Delegated Execution on Three Chains}

To trigger the malicious \texttt{fallback()} handler, we sent a self-call from the victim on each chain.

\smallskip
\noindent\textbf{Chain A trigger.}
\vspace{-0.12in}
\begin{lstlisting}[language=text]
txHash = 0xbe2f31a4519c97332...3955c46c329b0fac4
\end{lstlisting}
\vspace{-0.12in}
% \smallskip
\noindent\textbf{Chain B trigger (nonce=1).}
\vspace{-0.12in}
\begin{lstlisting}[language=text]
txHash = 0x3b955c7e7f6fc6e...b51fca0f72cb57954dc
\end{lstlisting}
\vspace{-0.12in}
% \smallskip
\noindent\textbf{Chain C trigger (nonce=2).}
\vspace{-0.12in}
\begin{lstlisting}[language=text]
txHash = 0x89ae46d7c6947...276b92535c1352fa3c35b
\end{lstlisting}
\vspace{-0.12in}
Execution traces confirmed that:
\begin{itemize}
  \item The victim’s call routed into \textcolor{teal}{adr.18bc},
  \item All ETH (minus gas residual) was swept to \textcolor{teal}{adr.79c8},
  \item All 2{,}000 tokens were transferred to the attacker,
  \item The fallback completed without revert.
\end{itemize}

% \newpage

% \section{erc7702 documentation}

% 1. \url{https://eip7702.io/}

% 2. \url{https://eips.ethereum.org/EIPS/eip-7702}

% 3. \url{https://ethereum.org/en/roadmap/pectra/}

% 4. \url{https://www.certik.com/zh-CN/resources/blog/pectras-eip-7702-redefining-trust-assumptions-of-externally-owned-accounts}

% \section{referred article}

% 1. \url{https://decentralizedsecurity.es/eip-7702-ethereums-next-step-toward-a-more-flexible-account-model}

% \section{attack news}
% 1. \url{https://www.bitget.com/news/detail/12560604791489}

% 2. \url{https://cryptorank.io/news/feed/2c18b-wlfi-token-holders-lose-millions-in-new-ethereum-phishing-attack}

% 3. \url{https://www.thecoinrepublic.com/2025/05/25/ethereum-news-how-hackers-are-exploiting-eip-7702-to-drain-wallets/}

% 4. \url{https://www.mexc.com/az-AZ/news/analysts-warn-of-1-5m-phishing-exploit-tied-to-ethereums-new-eip-7702/72493}

% 5. \url{https://www.cryptopolitan.com/phishing-scammers-infect-eip-7702-pectra/}

\vspace{0.2em}
\begin{center}
%\colorbox{blue!8}{
\fbox{
\begin{minipage}{0.93\linewidth}
\small
\textbf{RQ1 Findings.} 
EIP-7702 adoption exhibits clear security risks: 
(\textit{i}) delegation activity grows rapidly across chains; 
(\textit{ii}) authorization flows concentrate heavily on a few high-volume delegator contracts; 
(\textit{iii}) many of these endpoints belong to adversarial or suspicious categories. 
\end{minipage}
}
\end{center}
\vspace{0.2em}

\section{Empirical On-chain Measurement}
\label{app:pilot_study}
To understand the practical security implications of delegated execution under EIP-7702, we perform a pilot empirical study based on real-world on-chain activity. Our dataset combines transactions collected from major EVM chains with public blockchain analytics sourced from Dune Analytics~\cite{dune_query_6158978}. In total, our dataset aggregates \textbf{over 150k} 7702-related authorization and execution transactions collected from major EVM chains, spanning \textbf{more than 26k} unique victim addresses and \textbf{hundreds} of distinct delegator and vault-style contracts. This dataset forms the basis of our measurement of adoption patterns (RQ1) and real-world impact (RQ2).

\subsection{RQ1: How is EIP-7702 Being Adopted?}
To assess whether EIP-7702 creates a practically exploitable attack surface, we perform a cross-chain empirical study focusing on three core aspects.

\noindent\textbf{RQ1.1: Temporal growth of 7702 delegations.}
We first examine the temporal dynamics of 0x04-type delegation calls across chains. Fig.~\ref{fig:growth_trend} shows the daily and cumulative volume of EIP-7702 authorizations. Daily activity exhibits pronounced volatility, including repeated spikes reaching \emph{millions} of delegations per day. These bursts are inconsistent with organic user behavior and instead suggest automated issuance, potentially through bot-controlled wallets or phishing kits that batch delegation requests.

In contrast, the cumulative curve grows smoothly and monotonically, surpassing \emph{tens of millions} of total delegations within months. This pattern provides early evidence that 7702 delegations are being executed at scale, in ways that substantially affect real users.

\begin{figure}[t]
\centering

% ---- 上图 ----
\subfigure[Frequency Distribution of Delegator Contracts in EIP-7702 Transactions]{%
    \includegraphics[width=0.98\linewidth]{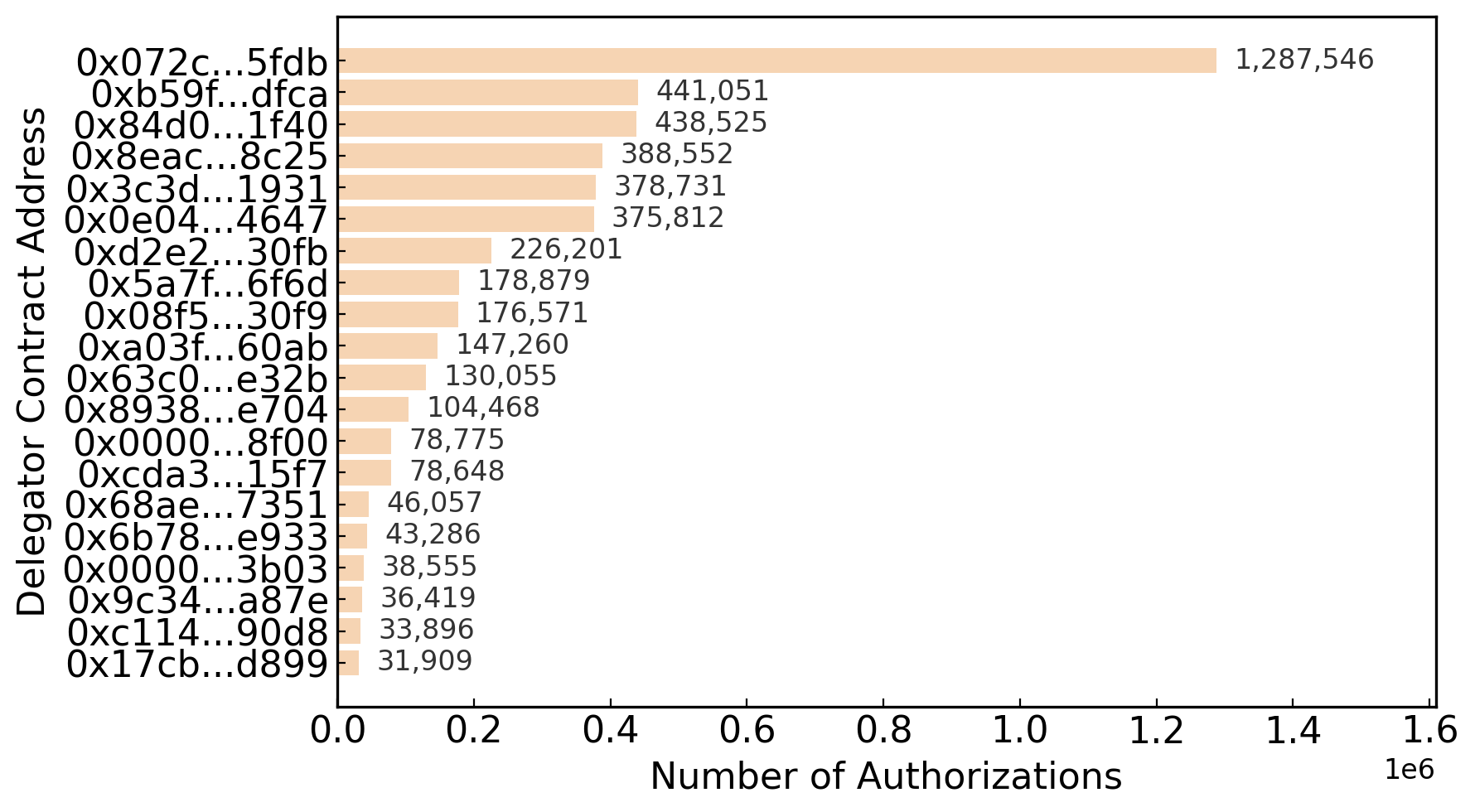}
    \centering
    \label{fig:delegator-frequency}
}

\vspace{0.8em}

% ---- 下图 ----
\subfigure[Delegator Contract Usage by Label Category]{%
    \includegraphics[width=0.9\linewidth]{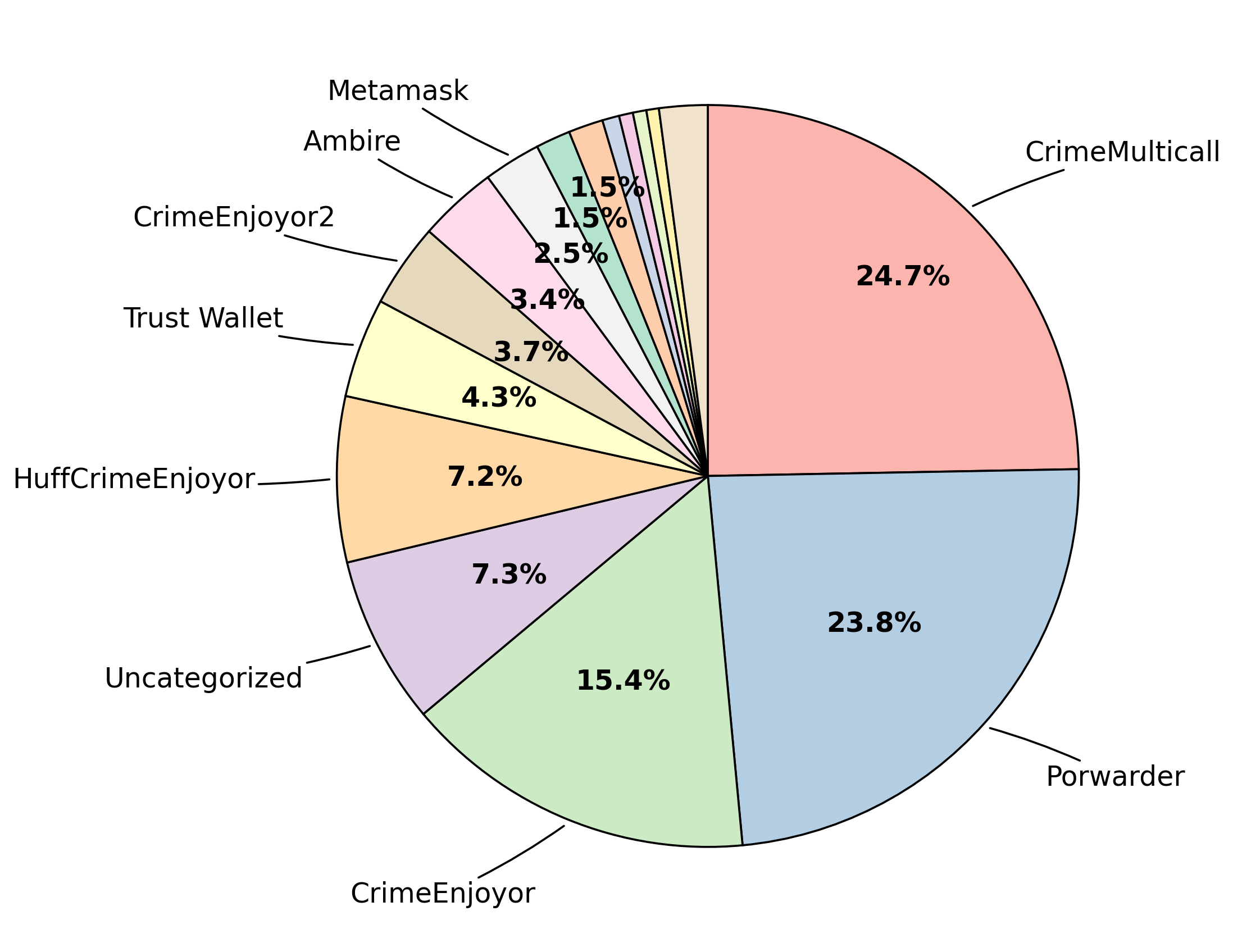}
    \centering
    \label{fig:delegator-category}
}
\caption{Distribution and categorization of delegator contracts invoked by EIP-7702 authorization transactions. }
\vspace{0.1in}
\end{figure}

\smallskip
\noindent\textbf{RQ1.2: Concentration of delegator contract endpoints.}
Next, we study which delegator contracts receive these authorization calls. Fig.~\ref{fig:delegator-frequency} reports the frequency distribution of the top contracts. 
The results reveal an extreme heavy-tailed structure: the most frequently used contract is authorized more than \emph{1.2\,M times}, nearly three times larger than the second-ranked address. A handful of contracts dominate the vast majority of delegation flows, while a long tail of lightly used contracts accounts for the remainder.

Such concentration signals systemic risk. Because an EIP-7702 delegation grants full operational authority over a user account until revoked, any vulnerability or malicious behavior in these high-volume endpoints can compromise a disproportionately large user population. Moreover, the sharp reuse patterns point to shared templates, e.g., wallet libraries or phishing toolchains, funneling users toward the same authorization targets.

\smallskip
\noindent\textbf{RQ1.3: Category-level characterization of delegator endpoints.}
We further label and cluster delegator contracts to understand the behavioral roles and ecosystem actors behind them. Fig.~\ref{fig:delegator-category} shows the category-level distribution, limited to families contributing at least 2\% of all delegations. Strikingly, crime-linked contract families (e.g., \textsc{CrimeMulticall}, \textsc{Porwarder}, \textsc{CrimeEnjoyor} and its variants) account for nearly half of all authorizations, far surpassing legitimate wallets or audited implementations such as \textsc{Metamask} or \textsc{Trust~Wallet}. This composition demonstrates that adversarial infrastructures are disproportionately responsible for shaping the real-world use of EIP-7702.

The prevalence of repeated contract variants further indicates that phishing groups and malicious operators reuse common templates, which substantially lowers the marginal cost of deploying new attacks. Additionally, a nontrivial fraction of contracts remains uncategorized, suggesting that our measurements may underestimate the true prevalence of adversarial sources.

\subsection{RQ2: What is the Impact of Malicious EIP-7702 Delegations?}
We seek to quantify the concrete economic consequences of 7702 abuse, e.g., how much victims lose, which delegator contracts are responsible, and how far the damage spreads across asset types such as ETH and NFTs.

\vspace{0.5em}
\begin{center}
%\colorbox{blue!8}{
\fbox{
\begin{minipage}{0.93\linewidth}
\small
\textbf{RQ2 Findings.} 
Malicious EIP\textendash7702 delegations cause substantial real-world harm: 
(\textit{i}) victim losses are heavily skewed, with several catastrophic multi-ETH liquidations; 
(\textit{ii}) a few high-impact delegator contracts account for most stolen value; 
(\textit{iii}) NFT assets are also drained at scale, with major collections losing hundreds of items. 
\end{minipage}}
\end{center}

\medskip
\noindent\textbf{RQ2.1: How are losses distributed across victims?}
We begin by quantifying the ETH stolen from users whose 7702 delegations were abused. Fig.~\ref{fig:top-eth-victims} reports the top ten victims ranked by total ETH loss. The distribution is highly skewed and exhibits strong heavy-tailed behavior. The largest victim suffered a catastrophic loss of over 3,300 ETH, more than five times the second-ranked address. Subsequent victims lost between 600--700 ETH, while a long tail of users lost 30--150 ETH each. These findings confirm that the attack surface created by EIP-7702 leads directly to high-magnitude monetary damage.

\medskip
\noindent\textbf{RQ2.2: Which delegator contracts are responsible for the largest aggregate losses?}
We next analyze losses aggregated by delegator contract, i.e., the entities empowered by 7702 delegations and subsequently used for theft. Fig.~\ref{fig:top-eth-delegators} ranks delegator contracts by the cumulative ETH stolen through them. The results reveal a small number of high-impact malicious endpoints. The top two delegators each caused more than 400 ETH of losses, closely followed by several others in the 300--380 ETH range.

The remarkably similar loss magnitudes across multiple delegators suggest stable, long-running phishing infrastructures rather than opportunistic one-off deployments. Attackers tend to repeatedly reuse the same delegator contracts, consistent with "scam-as-a-service" models that operate week to month timescales.

\medskip
\noindent\textbf{RQ2.3: How extensively are NFTs stolen?}
We examine NFT theft resulting from malicious delegations. Fig.~\ref{fig:top-nft-stolen} ranks NFT contracts by the number of unique items stolen. The top two collections each lost over 1{,}000 NFTs, while several others lost between 400--700 items.
Even contracts ranked near the bottom of the top-ten list experienced losses of 140--200 unique items.

These results confirm that 7702-based attacks are not limited to fungible tokens or ETH. Because NFT transfers do not implement allowance mechanisms and can be batched via multicall, a single malicious delegation enables attackers to drain \emph{all} NFTs held by a victim's account within one block, at negligible gas cost.

\end{document}